\documentclass[useAMS,usenatbib,usegraphicx]{mn2e}
\newcommand{\mcloud} {$M_{\rm cloud}$}      \newcommand{\Mcloud} {M_{\rm cloud}}
\newcommand{\rcloud} {$R_{\rm cloud}$}      \newcommand{\Rcloud} {R_{\rm cloud}}
\newcommand{\mudif}  {$\mu_{\rm dif}$}      \newcommand{\Mudif}  {\mu_{\rm dif}}
\newcommand{\mucol}  {$\mu_{\rm col}$}      \newcommand{\Mucol}  {\mu_{\rm col}}
\newcommand{\nh}     {$n_{\rm h}$}          \newcommand{\Nh}     {n_{\rm h}}
\newcommand{\nc}     {$n_{\rm c}$}          \newcommand{\Nc}     {n_{\rm c}}
\newcommand{\ndif}   {$n_{\rm dif}$}        \newcommand{\Ndif}   {n_{\rm dif}}
\newcommand{\ncol}   {$n_{\rm col}$}        \newcommand{\Ncol}   {n_{\rm col}}
\newcommand{\fracd}  {$F_{\rm dif}$}        \newcommand{\Fracd}  {F_{\rm dif}}
\newcommand{\fdif}   {$f_{\rm dif}$}        \newcommand{\Fdif}   {f_{\rm dif}}
\newcommand{\fcol}   {$f_{\rm col}$}        \newcommand{\Fcol}   {f_{\rm col}}
\newcommand{\mdif}   {$M_{\rm dif}$}        \newcommand{\Mdif}   {M_{\rm dif}}
\newcommand{\mcol}   {$M_{\rm col}$}        \newcommand{\Mcol}   {M_{\rm col}}
\newcommand{\tdif}   {$T_{\rm dif}$}        \newcommand{\Tdif}   {T_{\rm dif}}
\newcommand{\tcol}   {$T_{\rm col}$}        \newcommand{\Tcol}   {T_{\rm col}}
\newcommand{\tstop}  {$t_{\rm stop}$}       \newcommand{\Tstop}  {t_{\rm stop}}
\newcommand{\rstop}  {$R_{\rm stop}$}       \newcommand{\Rstop}  {R_{\rm stop}}
\newcommand{\tev}    {$t_{\rm ev}$}         \newcommand{\Tev}    {t_{\rm ev}}
\newcommand{\rev}    {$R_{\rm ev}$}         
\newcommand{\mev}    {$M_{\rm ev}$}         \newcommand{\Mev}    {M_{\rm ev}}
\newcommand{\msw}    {$M_{\rm sw}$}         \newcommand{\Msw}    {M_{\rm sw}}
\newcommand{\msnpl}  {$M_{\rm snpl}$}       \newcommand{\Msnpl}  {M_{\rm snpl}}
\newcommand{\edif}   {$E_{\rm dif}$}        \newcommand{\Edif}   {E_{\rm dif}}
\newcommand{\eth}    {$E_{\rm th}$}         \newcommand{\Eth}    {E_{\rm th}}
\newcommand{\ekin}   {$E_{\rm kin}$}        \newcommand{\Ekin}   {E_{\rm kin}}
\newcommand{\eturb}  {$e_{\rm turb}$}       \newcommand{\Eturb}  {e_{\rm turb}}
\newcommand{\vturb}  {$v_{\rm turb}$}       \newcommand{\Vturb}  {v_{\rm turb}}
  \newcommand{\Dkdis}  {\dot{e}_{\rm dis}}
   \newcommand{\Dkfb}   {\dot{e}_{\rm fb}}
\newcommand{\fexit}  {$f_{\rm exit}$}       \newcommand{\Fexit}  {f_{\rm exit}}
\newcommand{\dead}   {$\dot{E}_{\rm ad}$}   \newcommand{\Dead}   {\dot{E}_{\rm ad}}
\newcommand{\fcool}  {$f_{\rm cool}$}       \newcommand{\Fcool}  {f_{\rm cool}}
\newcommand{\fstar}  {$f_\star$}            \newcommand{\Fstar}  {f_\star}
\newcommand{\mstsn}  {$M_{\star{\rm ,sn}}$} \newcommand{\Mstsn}  {M_{\star{\rm ,sn}}}
\newcommand{\tlife}  {$t_{\rm life}$}       \newcommand{\Tlife}  {t_{\rm life}}
         \newcommand{\Nsn}    {N_{\rm sn}}
\newcommand{\rsn}    {$R_{\rm sn}$}         \newcommand{\Rsn}    {R_{\rm sn}}
\newcommand{\tpds}   {$t_{\rm pds}$}        \newcommand{\Tpds}   {t_{\rm pds}}
\newcommand{\rpds}   {$R_{\rm pds}$}        \newcommand{\Rpds}   {R_{\rm pds}}
\newcommand{\vpds}   {$v_{\rm pds}$}        \newcommand{\Vpds}   {v_{\rm pds}}
       \newcommand{\Tcool}  {t_{\rm cool}}
\newcommand{\dmcool} {$\dot{M}_{\rm cool}$} \newcommand{\Dmcool} {\dot{M}_{\rm cool}}
\newcommand{\dmsnpl} {$\dot{M}_{\rm snpl}$} \newcommand{\Dmsnpl} {\dot{M}_{\rm snpl}}
\newcommand{\dmsw}   {$\dot{M}_{\rm sw}$}   \newcommand{\Dmsw}   {\dot{M}_{\rm sw}}
\newcommand{\dmev}   {$\dot{M}_{\rm ev}$}   \newcommand{\Dmev}   {\dot{M}_{\rm ev}}
   
\newcommand{\decool} {$\dot{E}_{\rm cool}$} \newcommand{\Decool} {\dot{E}_{\rm cool}}
\newcommand{\desnpl} {$\dot{E}_{\rm snpl}$} \newcommand{\Desnpl} {\dot{E}_{\rm snpl}}
\newcommand{\defb}   {$\dot{E}_{\rm fb}$}   \newcommand{\Defb}   {\dot{E}_{\rm fb}}
\newcommand{\tdestr} {$t_{\rm destr}$}

\newcommand{\flost}{$f_{\rm lost}$}
\newcommand{\esn}{$E_{51}$}
\newcommand{\lmech}{$L_{38}$}
\newcommand{\msun}{${\rm M}_\odot$}

\newcommand{\be}{\begin{equation}}
\newcommand{\ee}{\end{equation}}
\newcommand{\fr}{_{\rm frag}}
\newcommand{\circa}{$\sim$}
\newcommand{\cmt}{cm$^{-3}$}
\newcommand{\kms}{km s$^{-1}$}

\title[On the destruction of star-forming clouds]
{On the destruction of star-forming clouds}
\author[P. Monaco]{Pierluigi Monaco\\
Dipartimento di Astronomia, Universit\`a di Trieste, 
via Tiepolo 11, 34131 Trieste, Italy - email: monaco@ts.astro.it}

\begin{document}

\date{Accepted ... Received ...}

\pagerange{\pageref{firstpage}--\pageref{lastpage}} \pubyear{2004}

\maketitle

\label{firstpage}

\begin{abstract}

Type II supernovae (SNe), probably the most important contributors to
stellar feedback in galaxy formation, explode within the very dense
star-forming clouds, where the injected energy is most easily radiated
away.  The efficiency of type II SNe in injecting energy into the
interstellar medium (ISM) and in re-heating a fraction of the original
star-forming cloud is estimated with the aid of a two-phase model for
the ISM of the cloud.  We argue that when SNe explode the star-forming
cloud has already been partially destroyed by ionizing light and winds
from massive stars.  SN remnants (SNRs) will first cause the collapse
of most of the cloud gas into cold fragments, until the diffuse hot
phase has a low enough density to make further radiative losses
negligible.  This is completed in \circa3 Myr, with a modest energy
loss of \circa5 per cent of the total budget.  We compute that a
fraction ranging from 5 to 30 per cent of the cloud is reheated to a
high temperature (from $10^5$ to $10^7$ K); these numbers are very
uncertain, due to the very complicated nature of the problem, and to
the uncertain role of thermal evaporation.  Small star-forming clouds,
less massive than \circa$10^4$ \msun, will be destroyed by a single
SN.  In all cases, a high fraction of the energy from type II SNe
($\ga$80 per cent for large clouds, smaller but still significant for
small clouds) will be available for heating the ISM.

\end{abstract}

\section{Introduction}

The formation of stars and galaxies, as well as the state of the ISM,
are regulated by the feedback processes related to the energy
injection into the ISM itself by stars through winds, UV photons and
SN explosions.  In particular, a satisfactory model of galaxy
formation requires that a significant fraction of the energy released
by massive stars and SNe is given to the ISM and eventually to the
hot, virialized gas component pervading the dark-matter halos.

The greatest part of the stellar energy budget is provided by type II
SNe, associated to short-lived massive stars.  The cosmological
community often restricts to this feedback source alone, thus
neglecting not only the contribution of UV and winds, that are
produced by a subset of the stars that die as SNe and whose energy
budget is less than the uncertainty in the energy of the single SNR,
but also the contribution of type Ia SNe, associated to less massive
stars.  This way, energy is injected where young stars reside, i.e. in
the star-forming regions, that are systematically the densest ISM
regions in a galaxy.  In this context the energy of SNe, that
propagates into the ISM through blast waves, is very easily radiated
away, giving rise to a low efficiency of energy injection.

The computation of the energy lost by a SNR while it gets out of a
star-forming cloud is not easy, due to a number of complications.  OB
stars are highly clustered, both within the galaxy (they reside in the
star-forming, molecular clouds) and within the star-forming cloud
itself (they reside in associations).  This way, most SNe explode in
the hot bubbles created by previous explosions, giving rise to
super-bubbles (SBs) more than isolated SNRs.  Besides, the
star-forming clouds are highly inhomogeneous, magnetized and dominated
by supersonic turbulence.

A further crucial point is that the first SNe explode from massive
progenitor stars that have already pre-heated the surrounding ISM by
UV light and winds.  This process has been addressed by many authors,
as for instance McKee, van Buren \& Lazareff (1984), McKee (1989),
Franco, Shore \& Tenorio-Tagle (1994), Williams \& McKee (1997),
Matzner (2002), Tan \& McKee (2004).  Despite all of the
uncertainties, some consensus is emerging on the fact that UV light
and winds from massive stars are able both to sustain the observed
level of turbulence in the star-forming clouds and to destroy them in
a time of order of 10Myr.  As a result, the bulk of SNe explode when
the cloud is already in an advanced state of destruction.

In a recent paper (Monaco 2004, hereafter paper I) we proposed a model
for feedback in galaxy formation in presence of a multi-phase medium.
In that model, the most difficult piece of astrophysics to address was
the efficiency of feedback in re-heating a fraction of the
star-forming cloud to some high temperature, together with the amount
of energy lost by SNe before completing the destruction of the
star-forming cloud.  The first two quantities (fraction and
temperature of re-heated matter) were left as free parameters, the
last one (energy loss) was assumed to be negligible.

In this paper we compute these quantities through a simple model for
the explosion of SNe in a pre-heated star-forming cloud.  The model
assumes that SNe explode inside a cloud composed by a two-phase medium
in pressure equilibrium, with a hot diffuse phase and a cold phase
fragmented into clouds.  Due to the complexity of the problem, we are
well aware that the model presented here is too naive either to
represent the full complexity of the process or to give accurate
predictions for the quantities involved.  However, the results give
insight on the physical processes in play and are informative enough
at a qualitative level to constrain the order of magnitude of the
quantities cited above; they can be used as a guide for future
numerical simulations of the destruction of star-forming clouds.

The paper is organized as follows.  Section 2 presents the assumed
initial conditions of the cloud when SNe start to explode, the model
for the two-phase ISM, the fate of SNRs propagating in the cloud, the
mass and energy flows within the components and the system of
equations used.  Section 3 presents the resulting predictions of the
model.  Section 4 discusses the case of small and dense star-forming
clouds, where only a few SNe per cloud explode, and Section 5 gives
the conclusions.

\section{The model}

\subsection{The state of the cloud at the first SN explosions}

Molecular clouds are dominated by supersonic turbulence (see, e.g.,
Solomon et al. 1987).  According to recent simulations (Mac Low et
al. 1998; Ostriker, Gammie \& Stone 1999; Mac Low 2003) supersonic
turbulence in a compressible fluid decays over a few crossing times.
This is true also in the case of magneto-hydrodynamical (MHD)
turbulence.  Besides, star formation takes place on longer
times-scales and with a low efficiency, so that to obtain in a
turbulent cloud a significant fraction of mass in stars
(i.e. efficiency of star formation), comparable to the observed value
ranging from 1 to 10 per cent (see, e.g., Carpenter 2000), it is
necessary to sustain turbulence.  Moreover, observations suggest that
star-formation should not last more than \circa10 Myr (see, e.g.,
Elmegreen 2002).

UV light and winds from massive stars are likely responsible for the
destruction of the star-forming clouds.  For instance, Matzner (2002)
argued that the expanding HII regions are the most likely drivers of
turbulence, and that the gradual photo-dissociation of H$_2$ and the
expulsion of re-heated material leads also to the destruction of the
molecular cloud, self-limiting the efficiency of star formation to the
observed value.  Requiring an equilibrium between the turbulence
driven by the expanding HII regions and that dissipated by the
turbulent cascade, and taking into account the rate at which
``blister'' HII regions heat the gas to a temperature in excess of
$10^4$ K and expel it, he computed that a cloud will be destroyed in a
time ranging from 10 to 30 Myr, the lower values being valid for the
largest molecular clouds.  The efficiency of star formation resulted
\circa10 per cent, nearly independent of cloud mass.  His arguments
are strictly valid for clouds with dynamical times shorter than the
mean ionizing lifetime of massive stars (for a Milky Way ISM this
amounts to $M>10^5$ \msun) and with escape velocities smaller then
twice the ionized sound speed ($>10$ \kms).

According to this author, SNe do not contribute significantly to the
destruction of the cloud.  In fact, the first SNe explode \circa3 Myr
after the formation of their progenitor stars and are indeed
associated to the same OB stars that produce the HII regions.  They
are so massive that their remnants can go directly to the snowplow
stage (see below) while still in the early free expansion stage; they
end up being confined within the HII region, after having radiated
away most of their energy.  Their contribution to the momentum of the
HII region is thus small.  However, this argument is correct only for
the first SN that explodes in each OB association.  A second explosion
within the same association would propagate into a rarefied hot
bubble; in this case the blast would reach the ionization front before
losing much energy. Moreover, many SNe come from smaller stars, that
explode later and are not necessarily associated to big HII regions.
As a consequence, only the energy of the very first SNe will be
completely lost; besides, energy injection by multiple SNe will become
important later than 3 Myr, when the process of cloud destruction is
already in an advanced state.

\begin{figure}
\centerline{\includegraphics[width=8.4cm]{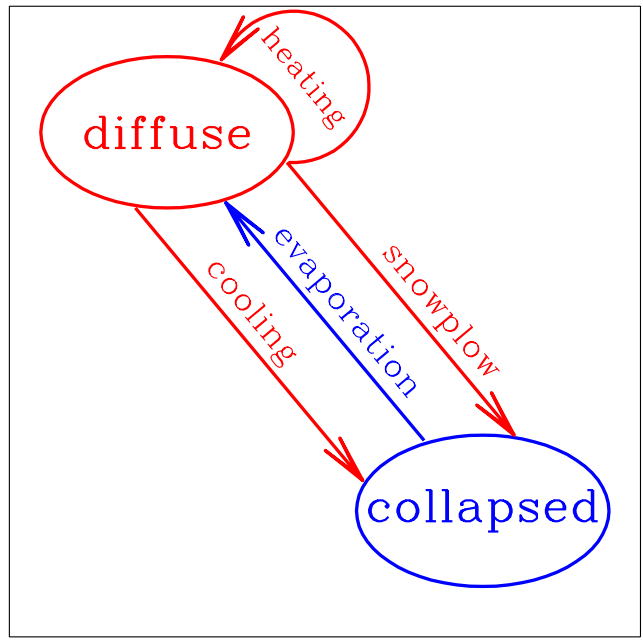}}
\caption{Mass flows in the model.}
\label{fig:fig1}
\end{figure}

While the precise origin of turbulence in star-forming clouds is still
to be demonstrated, the energetic input of OB stars is likely to have
a dramatic role in the destruction of the molecular, star-forming
clouds.  We thus deem it realistic to assume that when the first SNe
explode the cloud is composed of two phases, a hot diffuse one, heated
up by HII regions, and a cold collapsed one, fragmented into cloudlets
with a given mass spectrum.  Pressure equilibrium between the two
phases is assumed; this is justified by the finding of rough
isobaricity in simulations of turbulent ISM (Kritsuk \& Norman 2002;
see Mac Low 2002 and Vazquez-Semadeni 2002 for reviews).  The
expanding SNRs act in shaping the ISM within the cloud as follows
(Figure 1): blasts in the adiabatic stage heat the diffuse phase,
while in the snowplow stage (see below) they collapse it.
Thermo-evaporation of cold clouds within the expanding blasts
transfers mass from the collapsed to the diffuse phase.  Radiative
cooling transfers mass from the diffuse to the collapsed phase.

The initial conditions of the cloud are specified through its mass
\mcloud\ and initial radius \rcloud.  Paper I shows that these
quantities are related to the state of the ISM outside the cloud;
however, to keep the formalism simple we avoid making this connection
explicit.  In any case we assume that the external ISM is two-phase as
well, with densities of ``cold'' and ``hot'' phases \nc\ and \nh\
respectively.  Moreover, we assume for simplicity that the cloud is
spherical.  Following the discussion given above, a significant
fraction of mass is put into the diffuse phase; the temperature of the
collapsed phase is kept fixed to 100 K as in paper I.  A fraction
\fstar\ of the cloud is assumed to be in stars at the beginning of the
calculation, which coincides with the instant at which the first SN
explodes.  Any further star formation is neglected.

\subsection{The two-phase medium within the cloud}

The state of the ISM, the expansion of SNRs and the mass and energy
flows are modeled in a similar though simpler way as in paper I.  Let
\mdif\ and \mcol\ be the mass in the diffuse and collapsed phases
(with $\Mdif+\Mcol = (1-\Fstar)\Mcloud$), $\bar{\rho}_{\rm dif}$ and
$\bar{\rho}_{\rm col}$ their average densities ($\bar{\rho} =
3M/4\pi\Rcloud^3$), \tdif\ and \tcol(=100 K) their temperatures,
\mudif\ and \mucol\ their mean molecular weights and \fdif\ and \fcol\
their filling factors (with $\Fdif+\Fcol=1$).  Moreover, let
$\Fracd=\Mdif/(\Mcol + \Mdif)$ be the fraction of gas in the diffuse
phase.  The assumption of pressure equilibrium implies $P_{\rm th}/k=
\Ndif\Tdif = \Ncol\Tcol$, from which it is easy to obtain:

\be \Fcol = \frac{1}{1+\frac{\Fracd}{1-\Fracd}\frac{\Mucol}{\Mudif}
\frac{\Tdif}{\Tcol}} \label{eq:filcold} \ee

\noindent
(see Equation 2 of paper I).  

The collapsed phase is assumed to be fragmented into cloudlets, whose
mass spectrum $N\fr$ is:

\be N\fr(m\fr)dm\fr = N_0 (m\fr/1\ {\rm M}_\odot)^{-2} dm\fr\, .
\label{eq:cldistr} \ee

\noindent
Here $m\fr$ is the fragment mass in \msun\ and $N_0$ a normalization
constant such that $\bar{\rho}_{\rm col} = \int N\fr m\fr dm\fr$.  At
variance with paper I we fix the exponent of the mass function to
$-2$, which is a natural value in the presence of turbulence (see,
e.g., Elmegreen 2001).  The mass function is truncated below to a
value $m_l$ set to 0.1 \msun\ (see paper I and the discussions below)
and above by requiring unit probability for the existence of at least
one fragment (roughly $m_u=\Mcloud/\ln(\Mcloud/m_l)$).  A typical
radius $a\fr$ in pc is assigned to each cloud through the relation:

\be m\fr = 0.104\, \Mucol\Ncol a\fr^3\ {\rm M}_\odot \, . \label{eq:fragmass} \ee

\noindent
In the following we will assume that the fragments are spherical.

Finally, we call \vturb\ the rms kinetic velocity of the ISM,
$\Eturb=2\Vturb^2$ the kinetic energy per unit mass and $P_{\rm kin}$
the resulting kinetic pressure.

\begin{table}
\begin{center}
\begin{tabular}{lcl}
\hline
\multicolumn{3}{c} {Evaporative stage ($t<\Tev$ and $t<\Tpds$)}\\
\hline

$R_s(t)$        &=& $179\ (E_{51}\Sigma)^{1/10}\ t_6^{3/5}\ {\rm pc}$ \\
$v_s(t)$        &=& $105\ (E_{51}\Sigma)^{1/10}\ t_6^{-2/5}\ {\rm km\ s}^{-1}$\\
$\bar{T}(t)$    &=& $1.66\times10^5\ (E_{51}\Sigma)^{1/5}\Mudif \ t_6^{-4/5}\ {\rm K}$\\
$\Msw(t)$       &=& $5.91\times10^5\ (E_{51}\Sigma)^{3/10}\Mudif\Ndif \ t_6^{9/5}\ {\rm M}_\odot$\\
$\Mev(t)$       &=& $1.34\times10^4\ E_{51}^{4/5}\Sigma^{-1/5}\ t_6^{4/5}\ {\rm M}_\odot$\\
\tev            &=& $2.28\times10^4\ E_{51}^{1/2}\Sigma^{-1/2}(\Mudif\Ndif)^{-1}\ {\rm yr}$\\
\rev            &=& $18.5\ E_{51}^{2/5}\Sigma^{-1/5}(\Mudif\Ndif)^{-3/5}\ {\rm pc}$\\
\mev(\tev)      &=& $651\ E_{51}^{6/5}\Sigma^{-3/5}(\Mudif\Ndif)^{-4/5}\ {\rm M}_\odot$\\
\tpds           &=& $4.86\times10^3\ (E_{51}\Sigma)^{3/22}\Mudif^{15/22}\Ndif^{-5/11}\ {\rm yr}$\\
\rpds           &=& $7.32\ (E_{51}\Sigma)^{2/11}\Mudif^{9/22}\Ndif^{-3/11}\ {\rm pc}$\\
\vpds           &=& $884\ (E_{51}\Sigma)^{-1/22}\Mudif^{-3/11}\Ndif^{2/11}\ {\rm km\ s}^{-1}$\\
\eth            &=& $0.55\ E_{51}\ 10^{51}$ erg\\
\ekin           &=& $0.45\ E_{51}\ 10^{51}$ erg\\

\hline
\multicolumn{3}{c} {Adiabatic stage ($\Tev<t<\Tpds$)}\\
\hline

$R_s(t)$        &=& $84.5\ (E_{51}/\Mudif\Ndif)^{1/5}\ t_6^{2/5}\ {\rm pc}$ \\
$v_s(t)$        &=& $33.1\ (E_{51}/\Mudif\Ndif)^{1/5}\ t_6^{-3/5}\ {\rm km\ s}^{-1}$\\
$\bar{T}(t)$    &=& $4.65\times10^4\ (E_{51}/\Ndif)^{2/5}\Mudif^{3/5}\ t_6^{-6/5}\ {\rm K}$\\
$\Msw(t)$       &=& $6.22\times10^4\ E_{51}^{3/5}(\Mudif\Ndif)^{2/5}\ t_6^{6/5}\ {\rm M}_\odot$\\
\tpds           &=& $1.27\times10^4\ E_{51}^{3/14}\Ndif^{-4/7}\Mudif^{9/28}\ {\rm yr}$\\
\rpds           &=& $14.7\ E_{51}^{2/7}\Ndif^{-3/7}\Mudif^{-1/14}\ {\rm pc}$\\
\vpds           &=& $455\ E_{51}^{1/14}\Ndif^{1/7}\Mudif^{-11/28}\ {\rm km\ s}^{-1}$\\
\eth            &=& $0.72\ E_{51}\ 10^{51}$ erg\\
\ekin           &=& $0.28\ E_{51}\ 10^{51}$ erg\\

\hline
\multicolumn{3}{c} {PDS stage $(t>\Tpds)$}\\
\hline

$R_s(t)$        &=& $\Rpds\ (4t/3\Tpds-1/3)^{3/10}$\\
$v_s(t)$        &=& $\Vpds\ (4t/3\Tpds-1/3)^{-7/10}$\\
\eth(t)$^\dagger$  &=& $\Eth(\Tpds)\ \{0.398 [1-(t/1.169t')^{14/5}]$\\
  &&$+0.602[(R_s/R')^{10}+1]^{-1/5}[(t/t')^4+1]^{-1/9}\}$\\
\ekin(t)        &=& $\Ekin(\Tpds)((t-\Tpds)/\Tpds)^{-1/2}$ \\
\msnpl(t)       &=& $\Msw(t)\ [1-\Eth(t)/\Eth(\Tpds)]$\\

\hline
\end{tabular}
\label{table:snr}
\caption{Evolution of SNRs in the evaporative, adiabatic and PDS
stages.  $R_s$, $v_s$, $\bar{T}$, $\Msw$, $\Mev$, \eth\ and \ekin\ are
respectively the radius, velocity, average temperature, swept mass,
evaporated mass, thermal and kinetic energy of the SNR at a generic
time $t$.  The subscripts ev and pds refer respectively to the end of
the evaporative stage and to the onset of the PDS stage.  Here $t_6$
is time in units of $10^6$ yr and $T_6$ is \tdif\ in units of $10^6$
K. All these quantities are valid for solar metallicity and the simple
cooling function of Cioffi et al. (1988) and paper I.  $^\dagger$:
$t'=\Tpds\exp(1)$, $R'=R_s(t')$ and the first term in parenthesis is
present only for $\Tpds<t<3.16\Tpds$ (see Cioffi et al. 1988).}
\end{center}\end{table}

\subsection{The fate of SNRs}

One star with mass $>8$ \msun\ is formed each \mstsn\ \msun\ of stars
(taken to be 120 \msun), so the total number of SNe is
$\Nsn=\Fstar\Mcloud/\Mstsn$.  The rate of SN explosions, \rsn, is then
computed as:

\be \Rsn = \frac{\Fstar \Mcloud}{\Mstsn \Tlife} \label{eq:rsn}\, , \ee

\noindent
where \tlife\ is the difference between the lifetimes of an 8 \msun\
star and the most massive star, assumed to be \tlife=27 Myr.  Any
time-dependence of \rsn\ is neglected. Each SNR injects $E_{51}\
10^{51}$ erg of energy into the ISM.  \esn\ is conservatively taken to
be unity in the following, however its value is very uncertain and
could be significantly higher.  We call $E_{\rm sn}(t)$ the total
energy injected by SNe at the time $t$.  We assume that SNe are
homogeneously distributed within the cloud; this influences the
estimate of the porosity of SNRs.  The effect of the spatial
clustering of OB stars will be commented later.

Following McKee \& Ostriker (1977) and paper I, the blasts are assumed
to expand into the more pervasive diffuse phase; the cold dense
fragments pierce the blast, that reforms soon after the passage.  We
will ignore the effect of the collapsed phase on the blast.  At
variance with paper I, we take into account the thermo-evaporation of
cold clouds inside the expanding blasts, following the approach of
McKee \& Ostriker (1977) and Ostriker \& McKee (1988); the saturation
of thermo-evaporation is not taken into account.  Assuming that the
SNRs are not heavily mass loaded and do not lose thermal energy in the
early free-expansion stage, the expansion of the SNRs is divided into
three stages.  For sake of clarity, we report the main properties of
the evolution of the SNRs in Table 1.  At the beginning
thermo-evaporation, which depends on the 5/2-nd power of the average
internal temperature, is very efficient, so that the interior hot gas
is dominated by the evaporated mass.  In this case the evolution of
the remnant is given by the evaporative solution shown in Table 1,
where:

\be \Sigma^{-1} = \frac{3\Fcol\phi}{\alpha^2}
\left\langle\frac{1}{a\fr^2}\right\rangle_m \, .
\label{eq:sigma} \ee

\noindent
In this equation the parameter $\phi$ relates the actual thermal
conduction to the Spitzer value (used to compute the rate of thermal
evaporation), and its value depends sensitively on how effectively the
magnetic fields quench thermal conduction.  Besides, spherical
fragments present, at fixed mass and density, the smallest contact
surface between the two phases, so any non-sphericity of the cloud
will increase the effect of thermal evaporation; this can be mimicked
by an increase of $\phi$.  Moreover, evaporation acts on the low-mass
tail of the fragment mass function, and thus depends on the value of
$m_l$; keeping $m_l$ fixed, the uncertainty can be absorbed into
$\phi$.  As a result this parameter is highly uncertain, and every
value below or around unity is equally likely.  The parameter $\alpha$
relates the blast speed to the average sound speed of the interior,
and is in this case $\alpha^2=8$.  Finally, the quantity $1/a\fr^2$ is
averaged over the mass distribution of clouds
(Equation~\ref{eq:cldistr}).  It is worth noticing that $\Sigma$ has
the dimension of a surface (pc$^2$), and that it diverges
($\Sigma^{-1}$ vanishes) if thermal conduction is quenched.

At later times the evaporated mass \mev\ becomes smaller than the
swept mass \msw.  From this moment we use the standard adiabatic
solution, given again in Table 1.  Eventually, the interior gas cools
and collapses into a thin cold shell that acts as a snowplow on the
ISM.  For the evolution in the Pressure-Driven Snowplow (PDS) stage we
use the analytical model proposed by Cioffi, McKee \& Bertschinger
(1988), given again in Table 1, that fits reasonably well their
detailed 1D hydrodynamical simulations (confirmed by Thornton et
al. 1998).

The PDS stage can be reached when the blast is still in the
evaporative regime. In this case we pass directly to the PDS solution
of Cioffi et al. (1988), as the drop in the density of the internal hot
gas is very likely to quench thermal evaporation.

The thermal energy of the hot interior gas in the evaporative and
adiabatic stages is respectively 55 and 72 per cent, the rest being
kinetic.  In the PDS stage the evolution of the thermal energy is
given by Equation 3.15 of Cioffi et al. (1988) (reported in Table 1),
while the kinetic energy is assumed to decay as
$((t-\Tpds)/\Tpds)^{-1/2}$; this follows from assuming that all the
mass and kinetic energy is in the expanding shell, so that $E_{\rm
kin} = M_{\rm sw}v_s^{2}/2$, with $M_{\rm sw} \propto R_s^3$,
$v_s\propto R_s/t$ and $R_s\propto t^{0.3}$; this is valid for $t\gg
\Tpds$.  The interior mass is assumed to collapse into the snowplow at
the same rate at which thermal energy is lost.

The SNRs stop expanding in the following cases: (i) remnants stall by
pressure confinement; this happens when the velocity of the blast
equates the largest between the kinetic and the thermal velocity of
the ISM.  (ii) The porosity of the blasts, defined as $Q=\Rsn \int_0^t
R_s^3(t)dt/\Rcloud^3$ (and computed considering all the evolutionary
stages of the SNRs), reaches unit values.  The stopping time \tstop\
of the blast thus corresponds to the earlier of these events.

The porosity given above is valid for a homogeneous distribution of
SNRs.  The non-uniform spatial distribution of OB stars will influence
the estimate of $Q$ if the radius \rstop\ of the SNRs is small or
comparable to the typical distance between associations.  This is not
the case; we have verified that in most cases the stopping radius of
the SNR \rstop\ is similar to the initial size of the cloud.  This
confirms the validity of the uniform distribution of SNRs as a first
approximation.

\subsection{Mass and energy budget}

The onset of a PDS is important not only for its effect on the
evolution of the blast but also for its effect on the diffuse phase,
which is shocked to high temperature in the evaporative or adiabatic
stage but compressed and cooled in the PDS stage.

The rate at which the diffuse phase is swept by the ISM is $\Dmsw=\Rsn
\Msw(\Tstop)$.  Each SNR causes the following mass flows: (i) some
collapsed mass \mev\ is evaporated to the diffuse phase; (ii) some
diffuse mass \msnpl\ is collapsed if PDS is reached.  Both quantities
are given in Table 1 and are always computed at the time \tstop.  In
particular, the evaporated mass is given by the time-dependent term of
Table 1 for \tstop$<$\tev, while it remains constant after \tev\ (or
\tpds\ whenever it is smaller).  The evaporation and PDS mass flows
are then:

\be \Dmev = \Rsn \Mev(\Tstop)\, , \label{eq:dmev} \ee
\be \Dmsnpl = \Rsn \Msnpl(\Tstop)\, . \label{eq:dmsnpl} \ee

Radiative cooling of the diffuse phase leads to a global decrease of
thermal energy; in this process the density peaks cool dramatically
and thus move to the collapsed phase, giving rise to a cooling mass
flow:

\be \Dmcool = \Fcool \frac{\Mdif}{\Tcool}\, . \label{eq:dmcool} \ee

\noindent
As noted in paper I, the fraction \fcool\ depends on the detailed
density structure of the diffuse phase, which is very difficult to
predict without full-blown MHD simulation; notably, thermal conduction
also influences it.  \fcool\ is thus left as a free parameter.  The
cooling time is:

\be \Tcool=\frac{3k\Tdif}{\Ndif \Lambda(\Tdif)}\, , \label{eq:tcool} \ee  

\noindent
and is computed using the cooling function $\Lambda(T)$ of Sutherland
\& Dopita (1993)\footnote{We warn the reader that the analytical
quantities given in Table I are computed using the very simple cooling
function suggested by Cioffi et al. (1988) and used in paper I.};
solar metallicity will be used throughout the paper.  To avoid
overcooling at the beginning of the integration a heating source is
assumed to be present, such as to balance cooling for the diffuse
phase present at the initial time.  This is justified by the presence
of the same UV photons responsible for the destruction of the cloud,
and has an effect only during the first stages of evolution.

The thermal energy of the diffuse phase \edif\ is lost by radiation at
a rate:

\be \Decool = \frac{\Edif}{\Tcool}\, . \label{eq:decool} \ee

\noindent
The diffuse phase gains energy 
%
%
by blasts at a rate:

\be \Defb = \Rsn \Eth(\Tstop) \, , \label{eq:defb} \ee

\noindent
This includes the energy both in the evaporated and the heated gas.
The rate of energy loss by snowplows is:

\be \Desnpl = \Dmsnpl \Tdif \frac{3k}{2\Mudif m_p}\, . \label{eq:desnpl} \ee

\noindent
where $k$ is the Boltzmann constant and $m_p$ the proton mass.

The kinetic energy per unit mass \eturb, measured in km$^2$ s$^{-2}$,
of the collapsed phase decays through turbulent cascade as suggested
by Mac Low (2002, 2003):

\be \Dkdis = 6\times 10^{-7} \Vturb^3 L_d^{-1}\ {\rm km^2\ s^{-2}\ yr}^{-1}\, . \label{eq:dkdis} \ee

\noindent
Here $L_d$ is the driving scale of the turbulence (in pc), taken to be
twice the final diameter of the SNR (Mac Low 2002; Matzner 2002), or
$4\Rstop$.  The kinetic energy is continually replenished, provided
that the coherent velocity field of the expanding SNRs is randomized
by overlapping blasts.  The fraction of kinetic energy that goes out
of the cloud as a coherent velocity field is estimated as the ratio
between the initial external area of the cloud (the one that contains
all the stars) and the sum of all the areas of active blasts:

\be \Fexit = \frac{1}{R^2_{\rm cloud,i}} \int_0^{\Tstop}R_s^2(t) dt\, .
\label{eq:fexit} \ee

\noindent
As for the porosity, the integral is computed taking into account the
evolution of $R_s(t)$ in all the stages up to \tstop.  Of course
\fexit\ is not allowed to exceed unity.  The kinetic energy used to
drive turbulence is then:

\be \Dkfb = 5.03\times 10^7\ (1-\Fexit)\frac{\Rsn \Ekin}{\Mdif+\Mcol}
\ {\rm km^2\ s^{-2}\ yr}^{-1}\, . \label{eq:dkfb} \ee

\noindent
Here $5.03\times 10^7=10^{51}\ {\rm erg}/(10^{5}\ {\rm cm})^2\, {\rm
M}_\odot$.

The thermal energy of the diffuse phase and the kinetic energy not
used to drive turbulence are available for driving a SB into the
external two-phase ISM with densities of ``hot'' and ``cold'' phases
\nh\ and \nc\ (these phases have the same mean molecular weights as
the diffuse and collapses ones).  The exact modeling of the SB is
beyond the interest of this paper, but the radius \rcloud\ of the
destroyed cloud will expand with the SB.  To follow this expansion we
use some results that are valid in the adiabatic SB solution (Weaver
et al. 1977; paper I).  The velocity of the SB is assumed to be:

\be v_{\rm sb}(t) = 89.5 \left( \frac{L_{38}}{\Mudif\Nh} \right)^{1/3} 
R_{\rm sb}^{-2/3}\ {\rm km\ s}^{-1}\, , \label{eq:vsb} \ee

\noindent
where $R_{\rm sb}$ is in pc and the mechanical luminosity \lmech, in
units of $10^{38}$ erg s$^{-1}$, is:

\be L_{38} = 3.16\times10^5 (\Eth+\Fexit \Ekin) \Rsn\, . \label{eq:lmech} \ee

\noindent
The total energy used to drive the SB will be $E_{\rm sb}=\int
L_{38}\, dt$.  Moreover, we identify the radius of the shocked wind as
the time-dependent radius of the cloud, \rcloud, and set it to
0.86$R_{\rm sb}$.  This way \rcloud\ expands at a velocity:

\be v_{\rm exp}(t) = 0.86\ v_{\rm sb}(\Rcloud/0.86)\, . \label{eq:vexp} \ee

The adiabatic expansion of the cloud leads to a further energy loss
term for the diffuse phase.  According to Weaver et al. (1977), a
fraction 6/11 of the injected energy is given to the shocked external
ISM.  The energy lost by adiabatic expansion is then simply set as:

\be \Dead = \frac{6}{11} \Defb\, . \label{eq:dead} \ee

\noindent
We warn the reader that the predicted \tdif\ depends sensitively on
how the \dead\ term is modeled.

From all the mass and energy flows listed above the following system
of equations can be written:

\be \left\{
\begin{array}{lcl}
\dot{M}_{\rm dif}  &=& - \Dmcool - \Dmsnpl + \Dmev \\
\dot{M}_{\rm col}  &=&   \Dmcool + \Dmsnpl - \Dmev \\
\dot{E}_{\rm dif}  &=& - \Decool - \Desnpl + \Defb - \Dead \\
\dot{e}_{\rm turb} &=& - \Dkdis  + \Dkfb \\
\end{array} \right. \label{eq:system} \ee

\noindent
This set of equations, together with Equation~\ref{eq:vexp} for
\rcloud, is integrated with a standard Runge-Kutta code (Press et
al. 1992).

\begin{figure*}
\centerline{
\includegraphics[width=8.4cm,height=7.5cm]{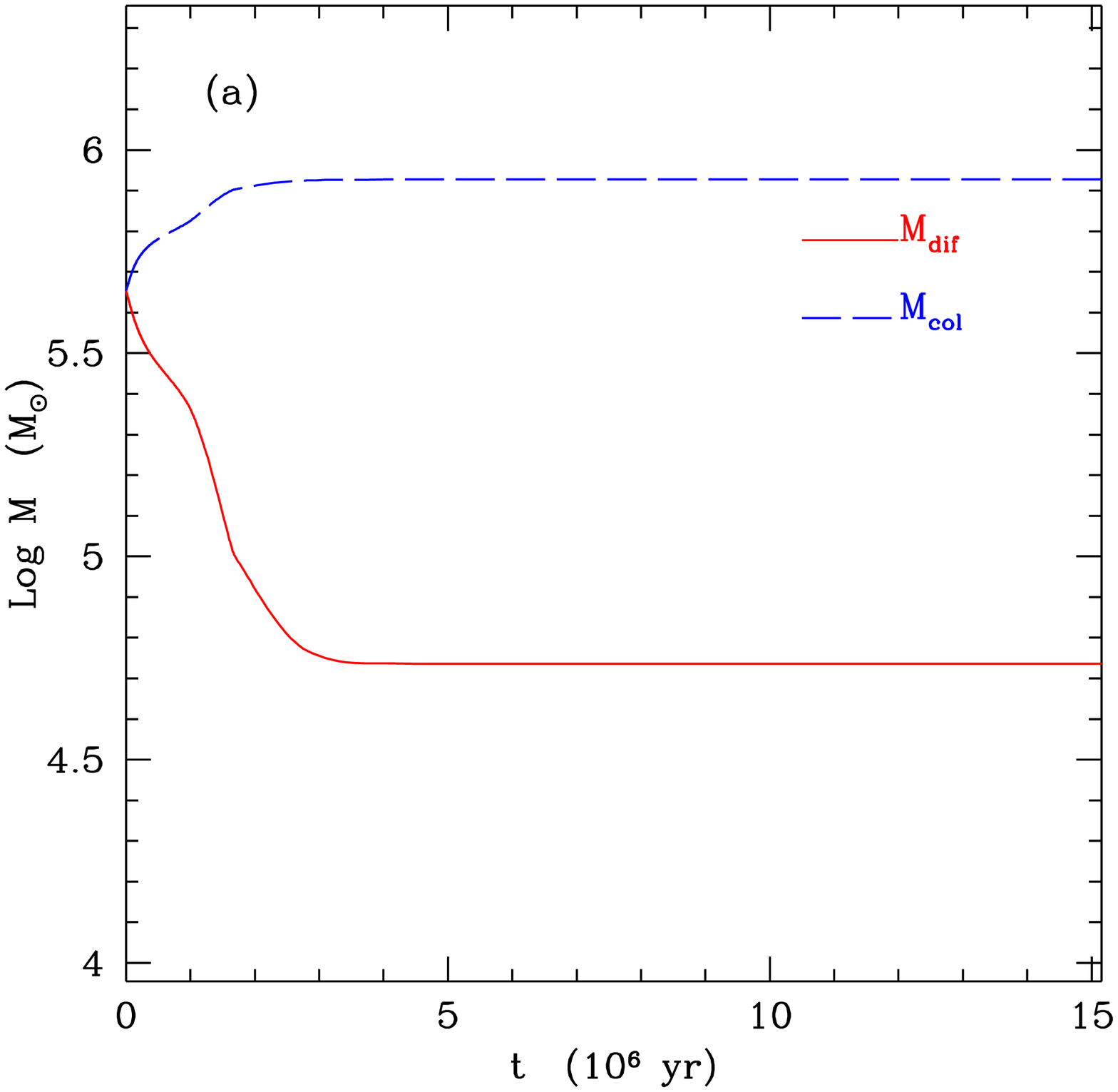}
\includegraphics[width=8.4cm,height=7.5cm]{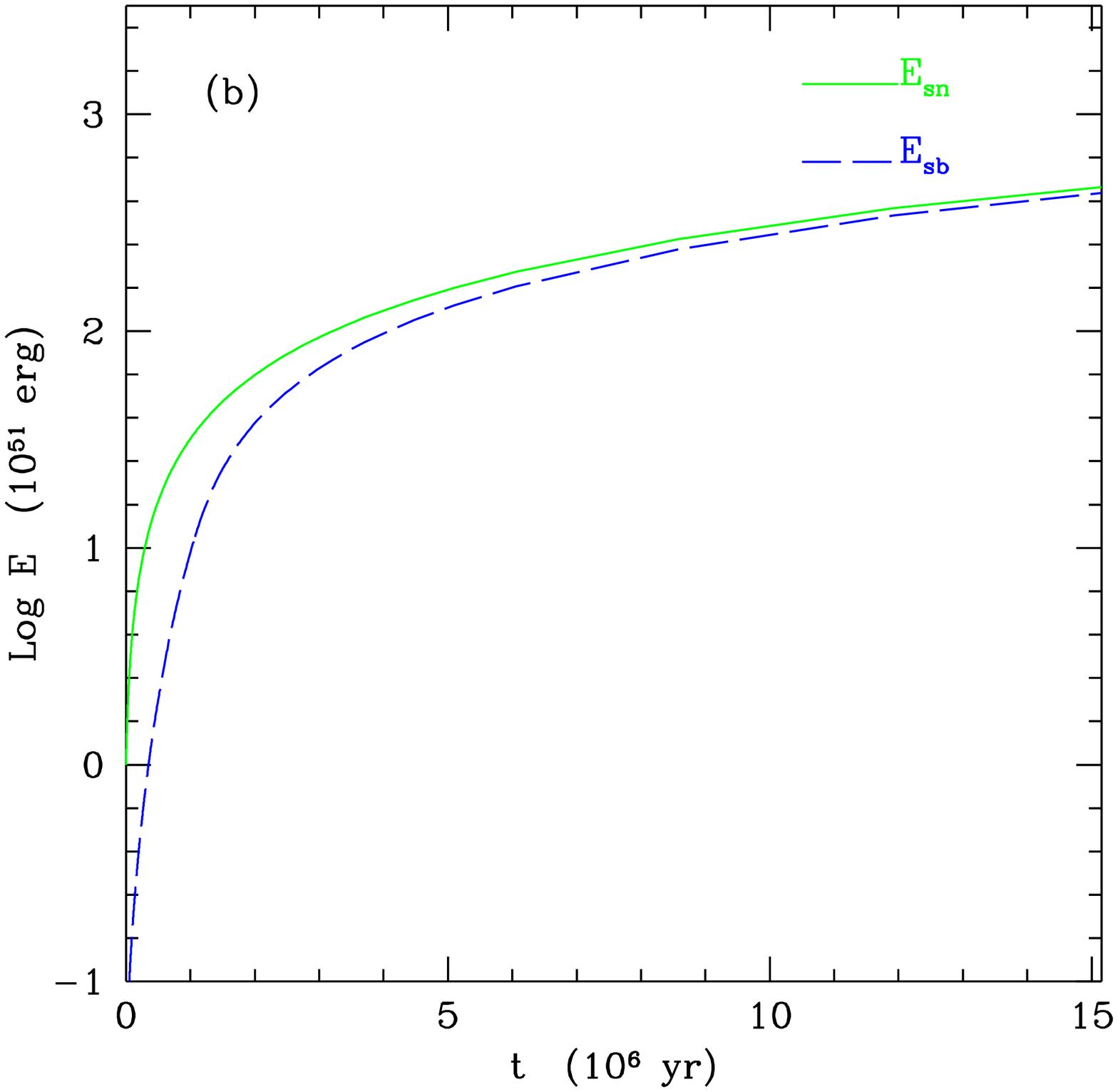}
}
\centerline{
\includegraphics[width=8.4cm,height=7.5cm]{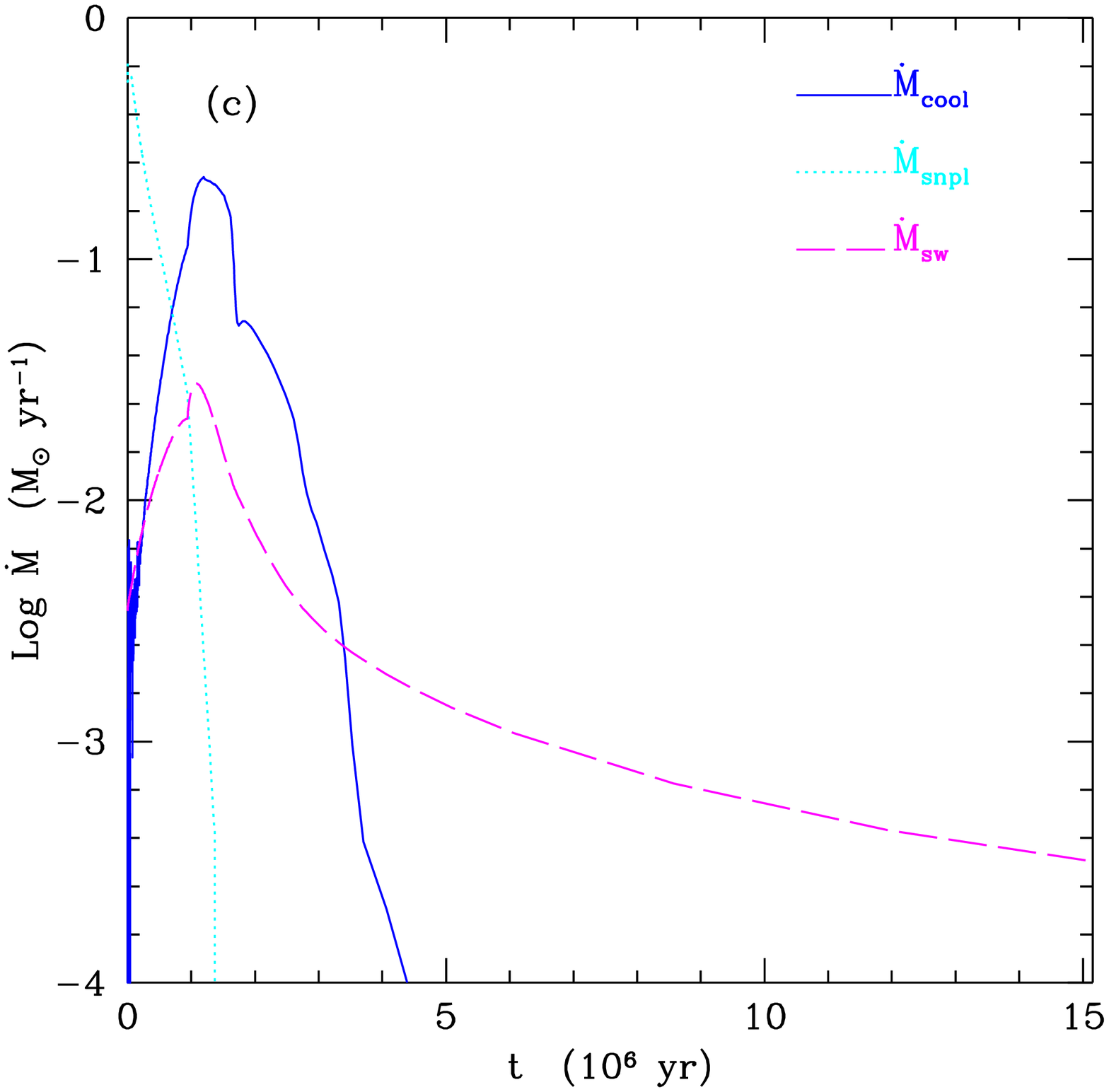}
\includegraphics[width=8.4cm,height=7.5cm]{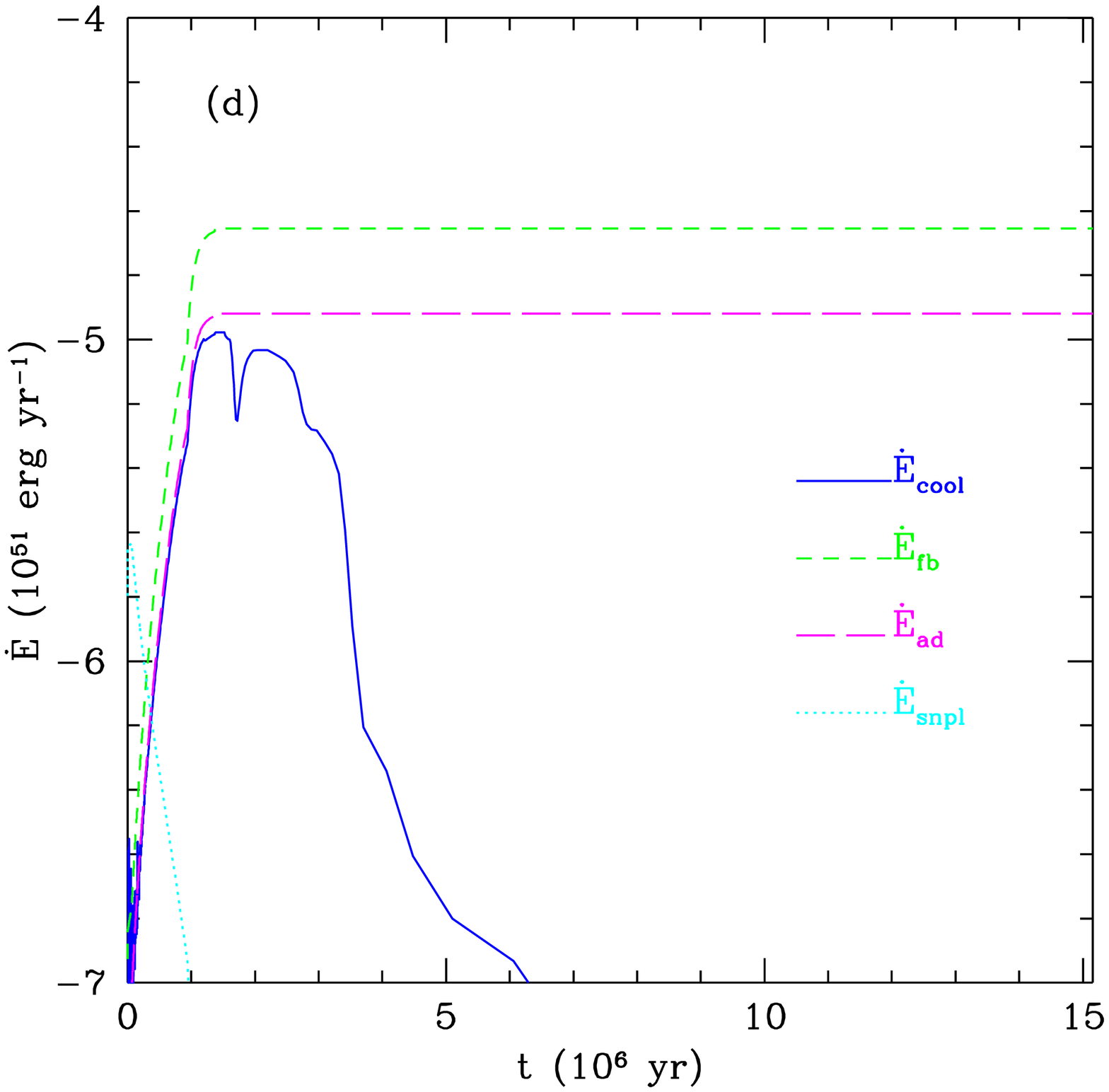}
}
\centerline{
\includegraphics[width=8.4cm,height=7.5cm]{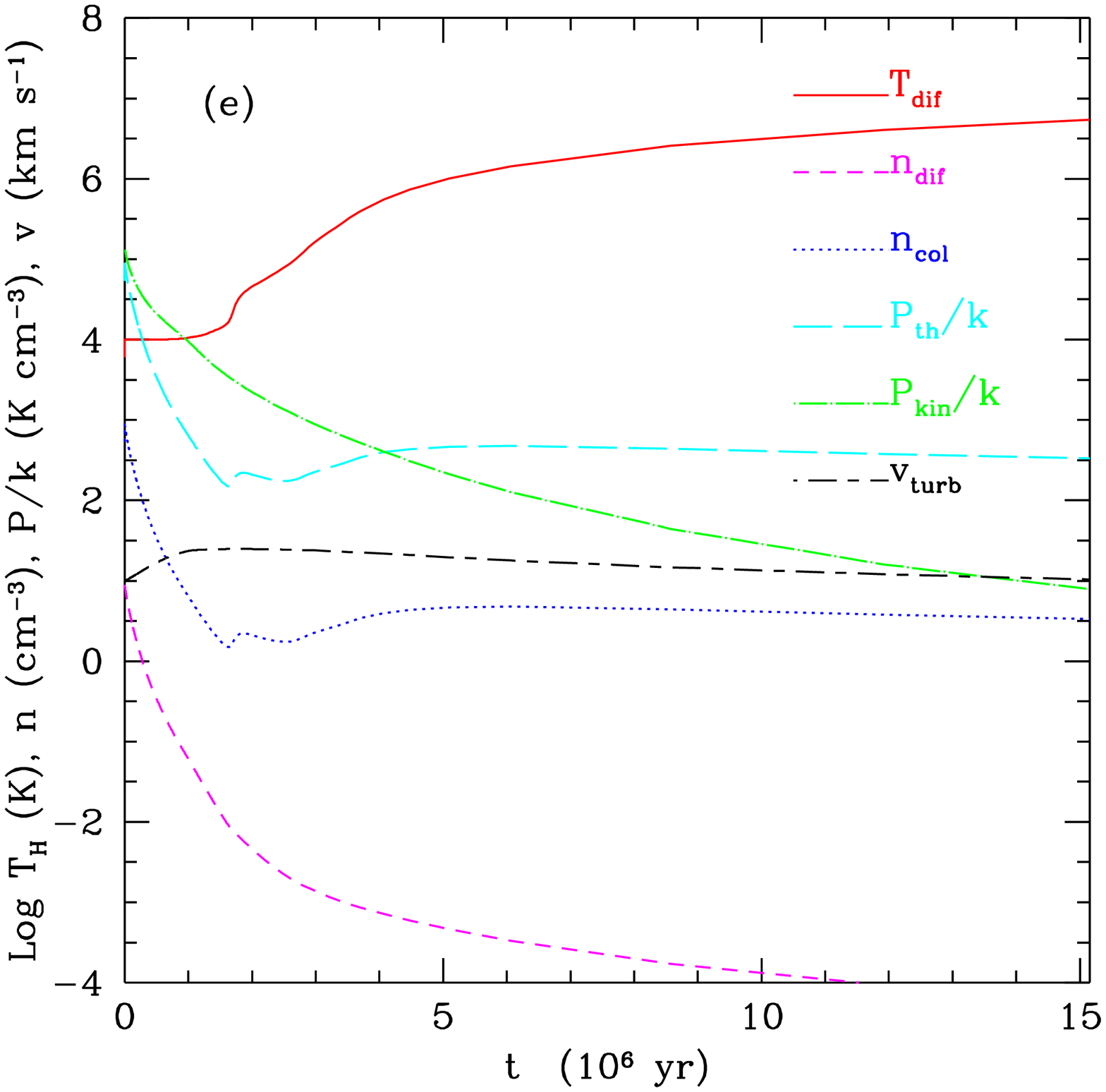}
\includegraphics[width=8.4cm,height=7.5cm]{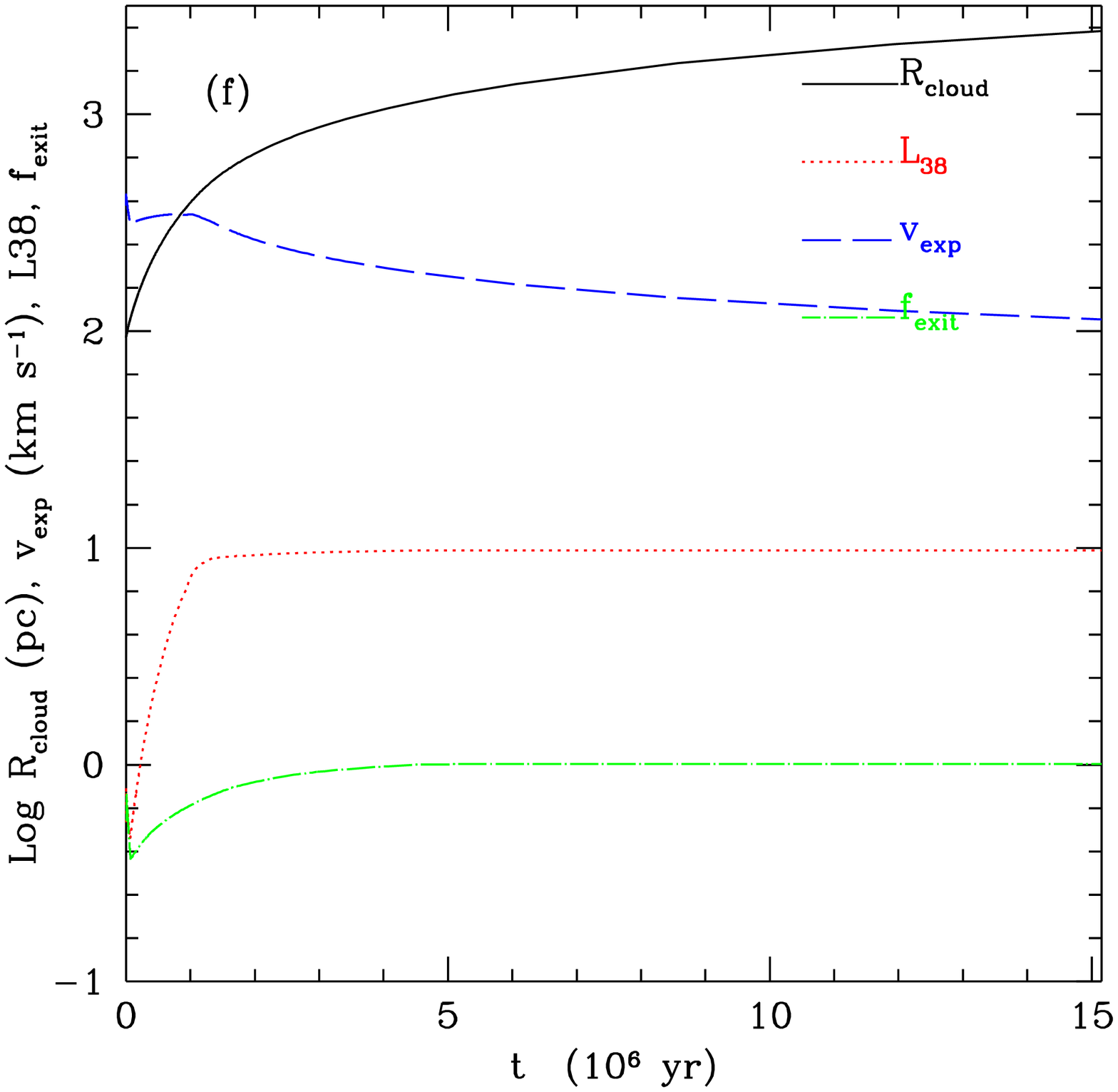}
}
\caption{Evolution of a cloud with \mcloud=10$^6$ \msun\ for the
reference choice of parameters given in the text. (a) Mass in the two
phases. (b) Energy used to drive the SB compared to that injected by
SNe. (c) Mass flows, including the rate at which mass is swept by
SNRs.  (d) Energy flows.  (e) State of the ISM.  (f) Mechanical
luminosity and expansion of the SB.}
\label{fig:std1e6}
\end{figure*}

\begin{figure*}
\centerline{
\includegraphics[width=8.4cm,height=7.5cm]{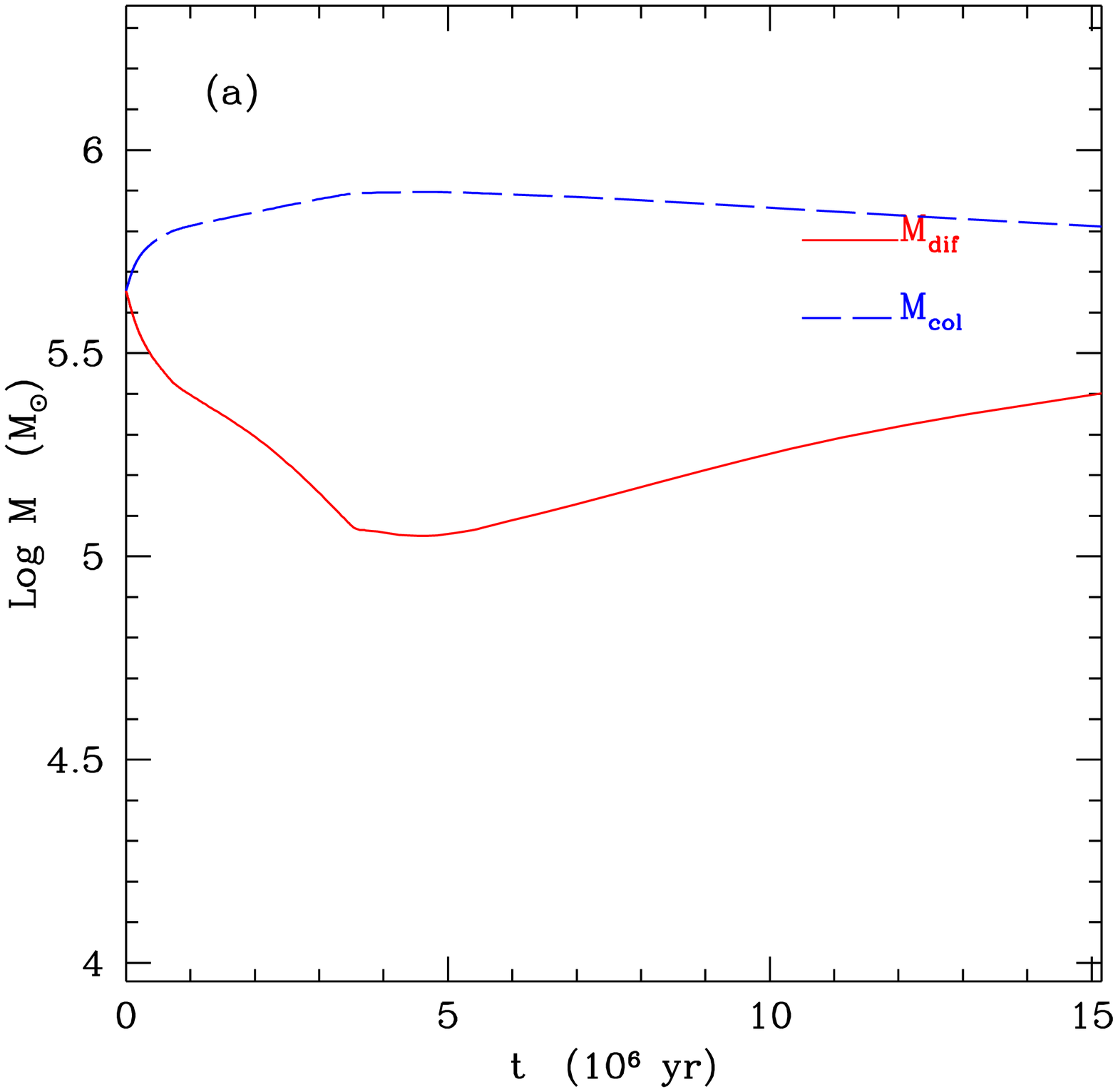}
\includegraphics[width=8.4cm,height=7.5cm]{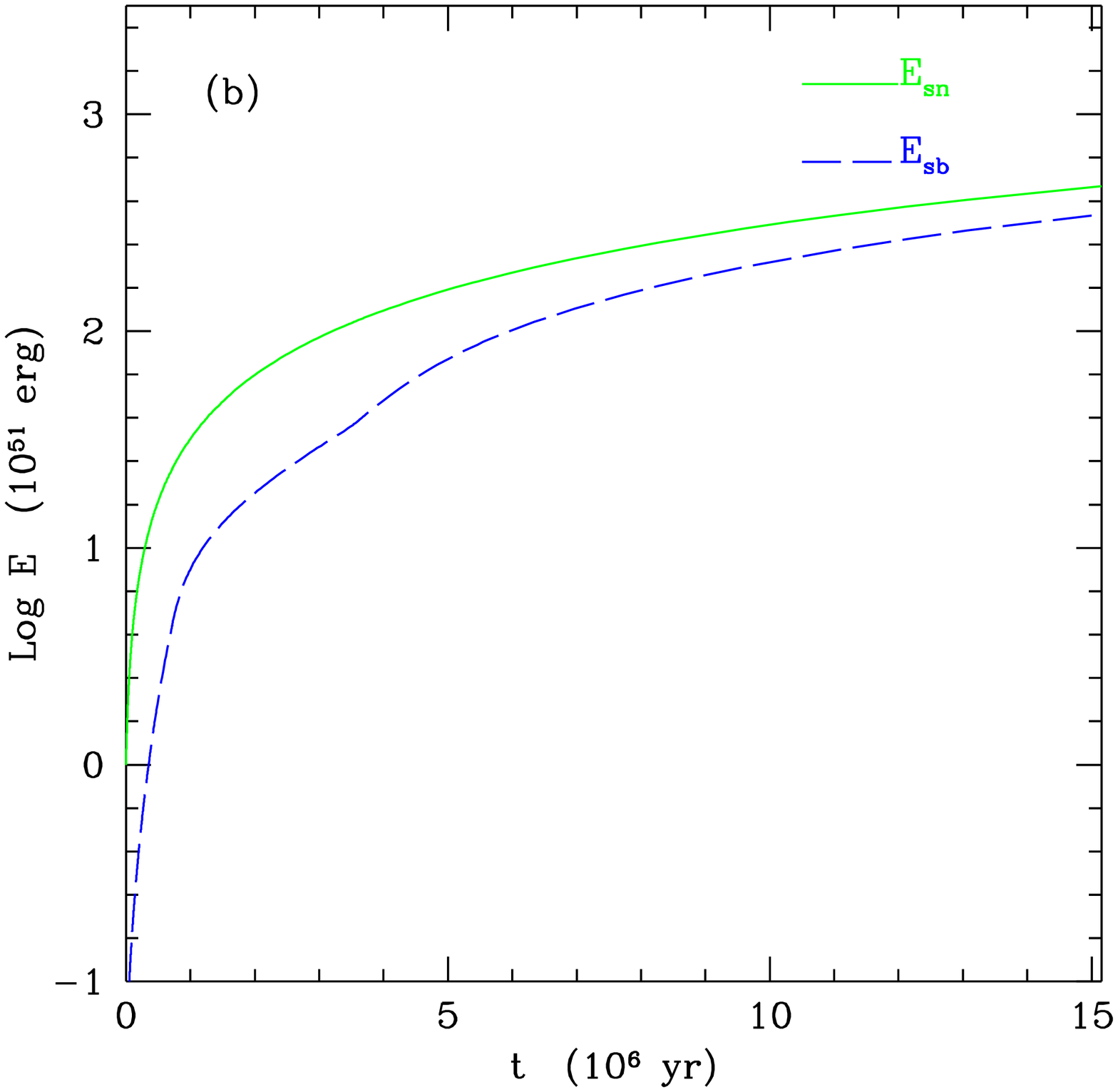}
}
\centerline{
\includegraphics[width=8.4cm,height=7.5cm]{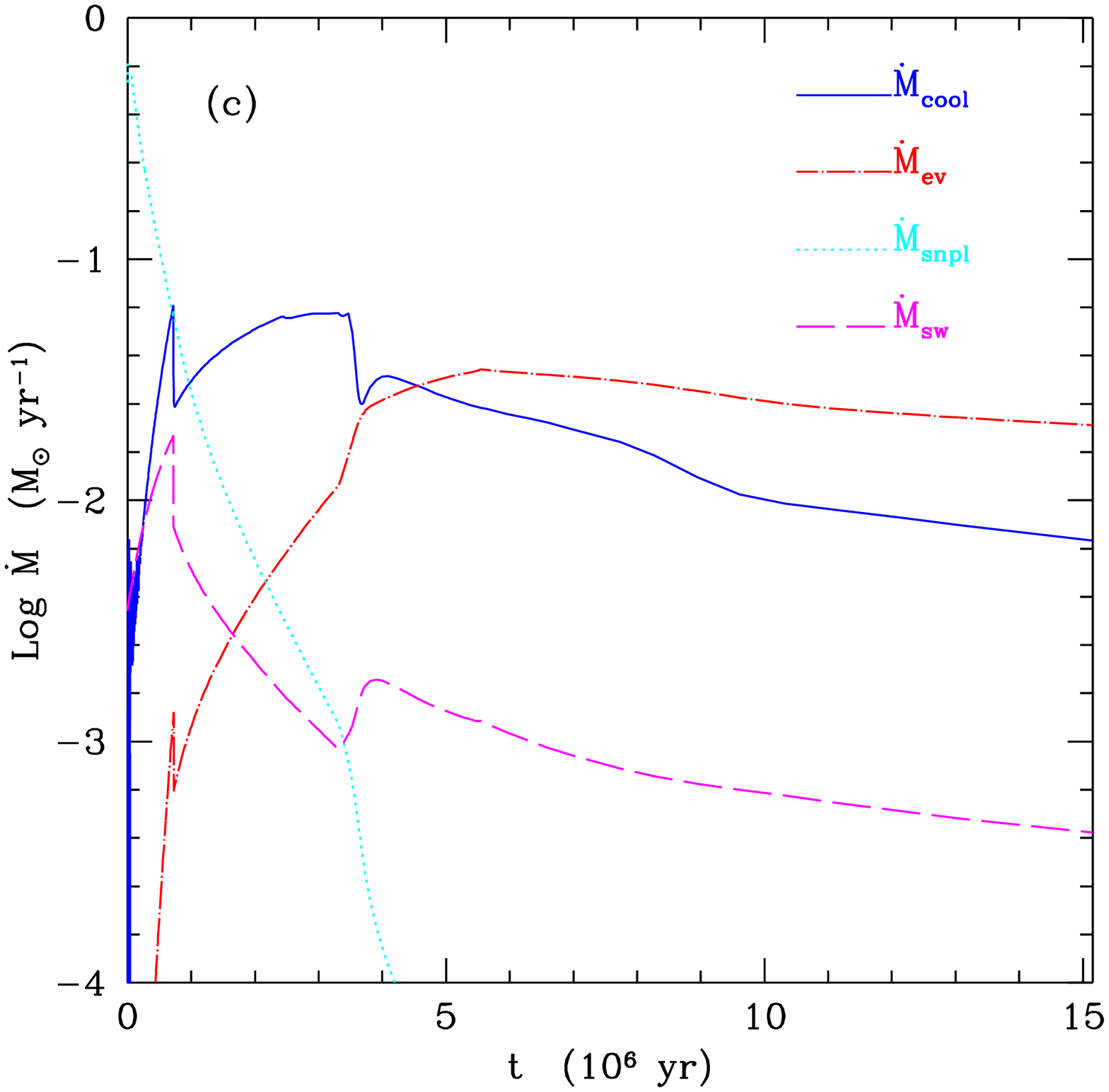}
\includegraphics[width=8.4cm,height=7.5cm]{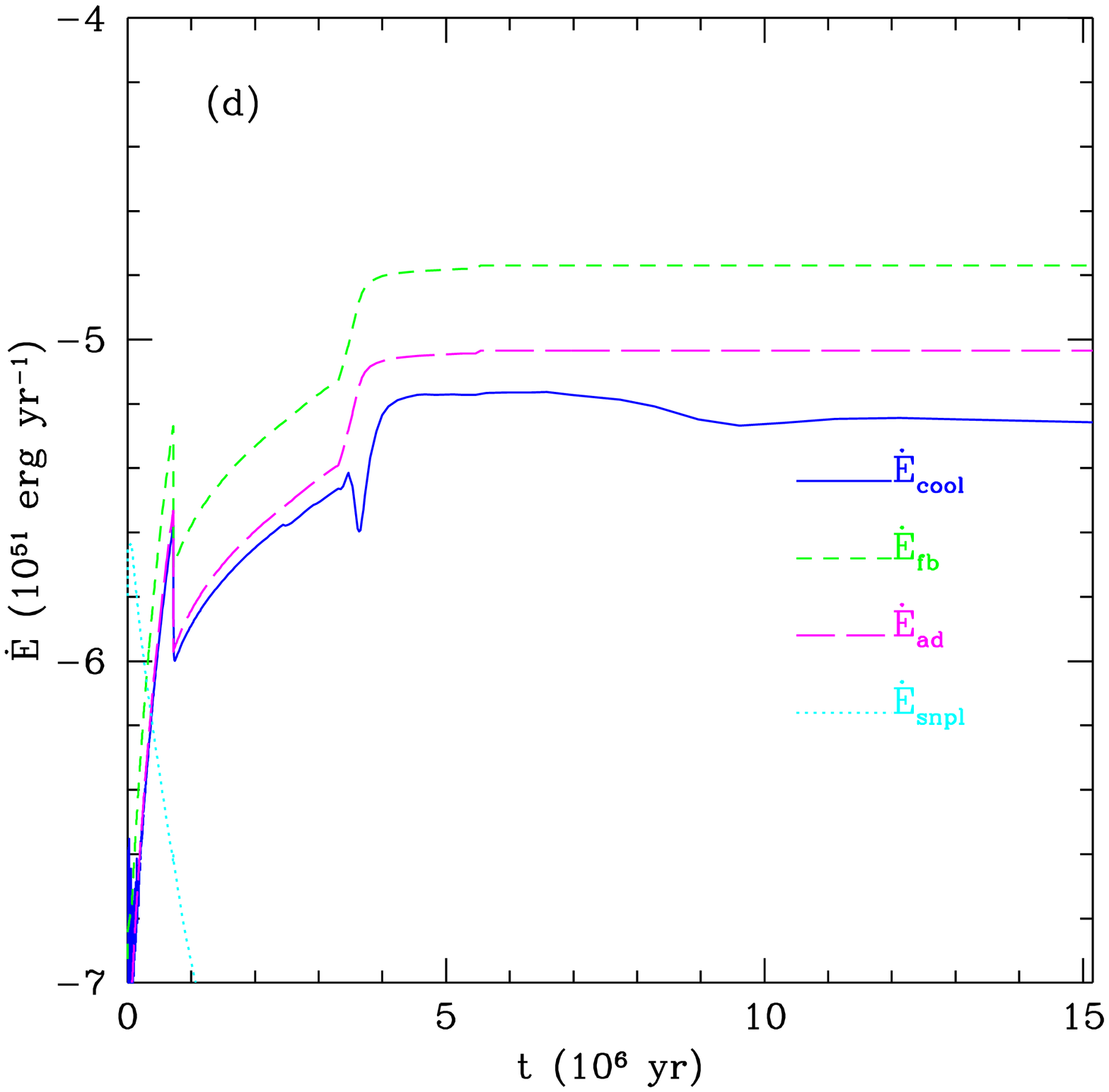}
}
\centerline{
\includegraphics[width=8.4cm,height=7.5cm]{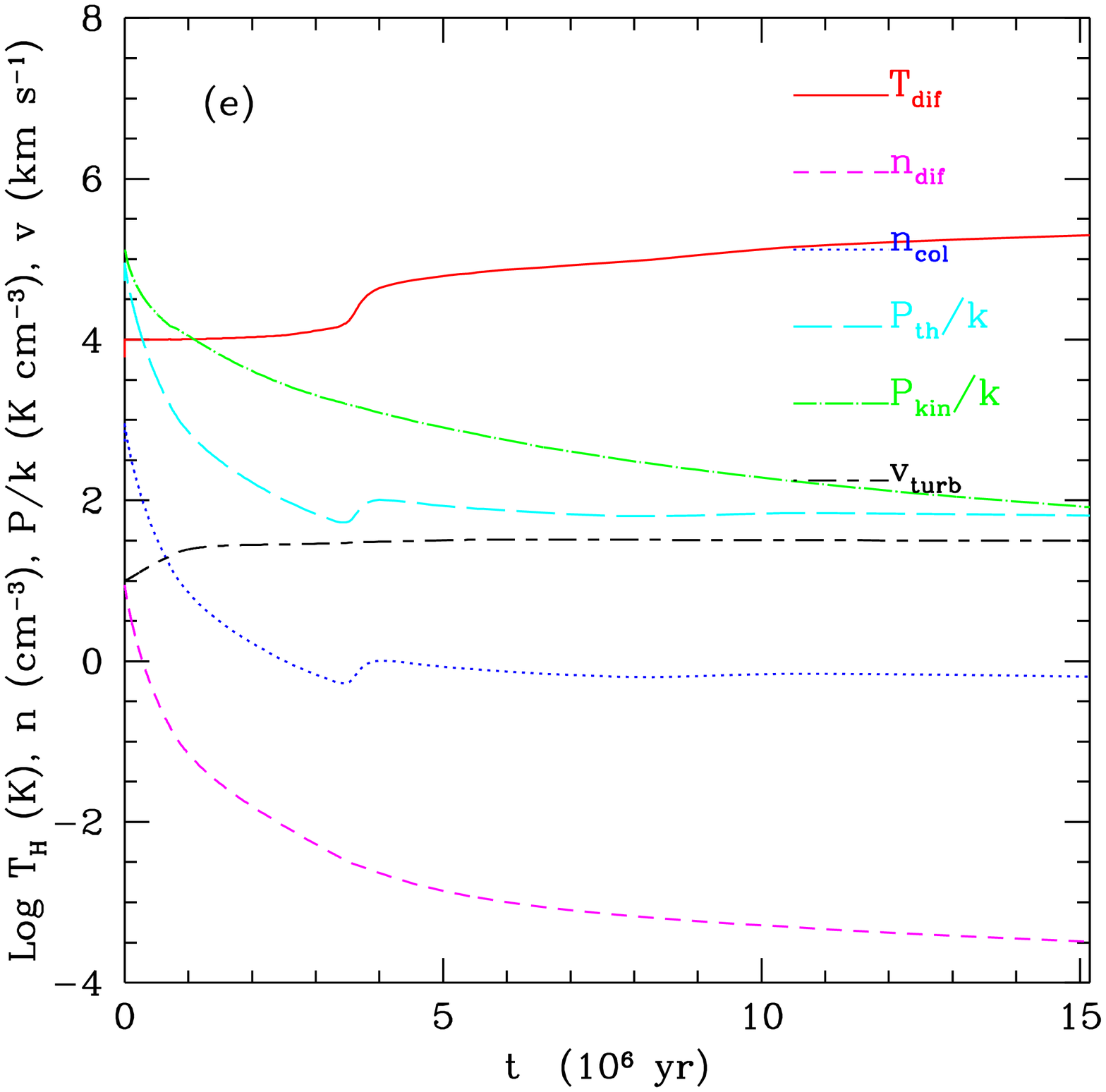}
\includegraphics[width=8.4cm,height=7.5cm]{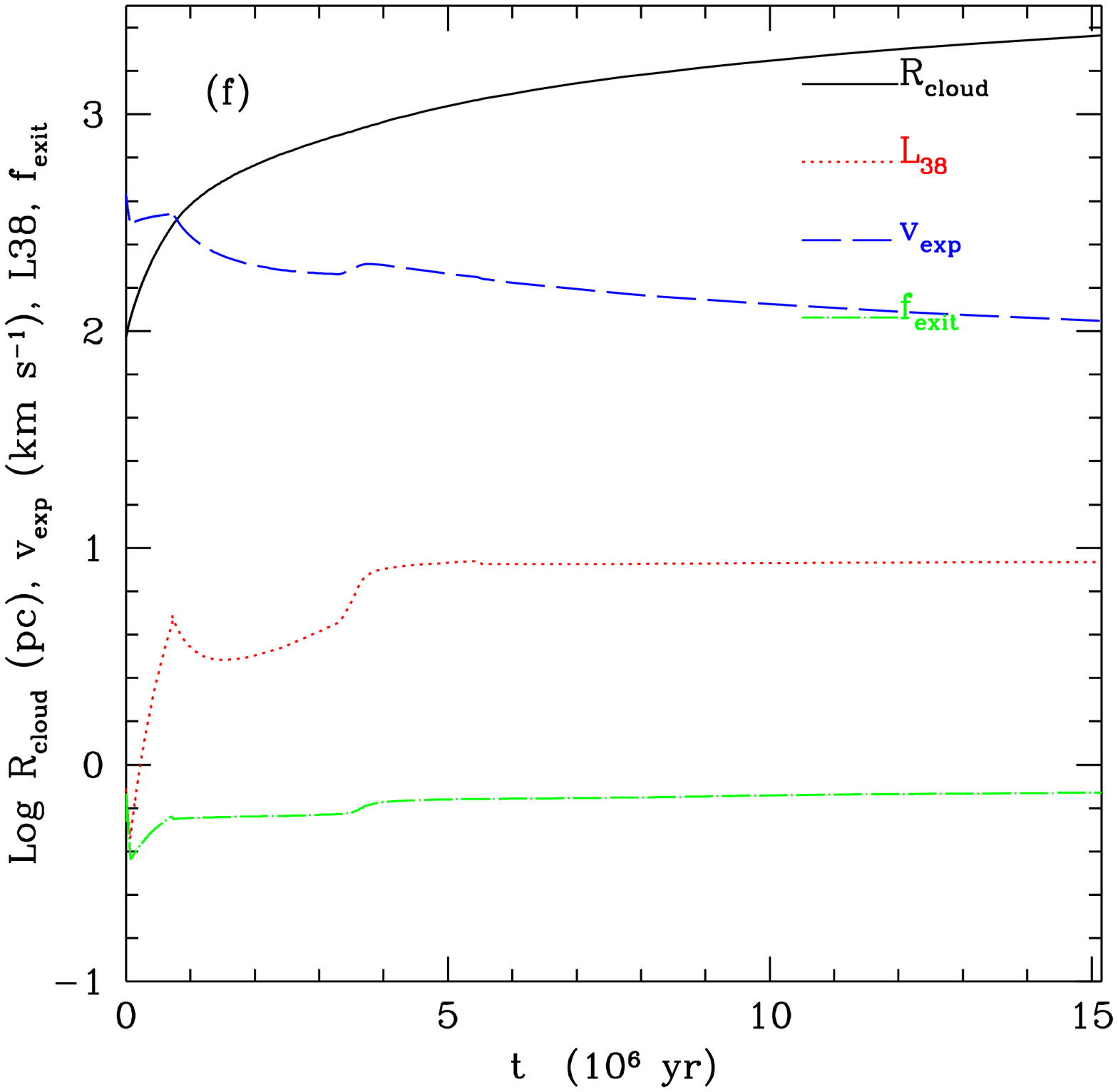}
}
\caption{The same as in Figure 2 for the example with $\phi=0.3$.}
\label{fig:phi0.3}
\end{figure*}

\section{Results}

The dynamical evolution of the system shows qualitative trends that
are present for a very wide range of initial conditions and
parameters.  These trends are well illustrated by the example shown in
Figure 2, relative to a cloud with \mcloud=10$^6$ \msun, with initial
conditions \mdif=\mcol=$(1-\Fstar)\Mcloud/2$, \tdif=10$^4$ K and
\vturb=10 \kms.  Parameter values are set to \esn=1, $\phi$=0 and
\fcool=0.1. The initial radius of the cloud is set using the same
Equation~\ref{eq:fragmass} with the external ``cold'' phase density
\nc\ in place of \ncol; for the external ISM we use as in paper I
\nc=10 and \nh=$10^{-3}$ \cmt, typical values for a galaxy disc.  The
figure shows the evolution of the masses of the two phases (\mdif\ and
\mcol), the energy released by SNe ($E_{\rm sn}$) versus the energy
used to drive the SB ($E_{\rm sb}$), the mass flows (\dmsw, \dmcool,
\dmsnpl\ and \dmev), the thermal energy flows (\decool, \desnpl,
\defb\ and \dead), the state of the ISM (\tdif, \ndif, \ncol, $P_{\rm
th}$, $P_{\rm kin}$ and \vturb) and the quantities characterizing the
expansion of the cloud within the SB (\rcloud, \lmech, $v_{\rm exp}$
and \fexit).

At the starting time the cloud is dense enough to allow SNRs to go
into the PDS stage before being stopped.  This creates a strong mass
flow from the diffuse to the collapsed phase, with a dramatic drop of
thermal pressure and densities of both phases.  This increase of the
mass of the collapsed phase does not imply a re-formation of the
molecular cloud, as the fragments will be spread out within the volume
of the expanding SB; they will later mix with the external cold phase.
In this early stage pressure is dominated by the kinetic contribution
due to the turbulent motion of the cold phase.  The porosity of SNRs
soon reaches unity values, that are maintained throughout the
evolution.  When the diffuse phase is significantly depleted the SNRs
merge before getting into the PDS stage; this happens in this example
after 1.3 Myr.  At this point energy is efficiently injected into the
diffuse phase, whose temperature starts to increase.  Once mechanical
heating overtakes the effective heating term introduced in the
equation, cooling becomes the dominant mass flow.  This induces a
further drop in the mass of the diffuse phase, which stabilizes after
\circa3 Myr to a constant value.  From that moment the fraction of
diffuse to collapsed mass and the density of the collapsed phase
remain constant, while the other quantities follow the expansion of
the cloud.  Notably, the temperature of the diffuse phase increases
gradually as a result of the efficient energy pumping.  The energy
used to drive the SB is significant after 1Myr, energy losses are
restricted to the first 3 Myr of evolution.  Finally, the turbulent
velocity \vturb\ remains between 10 and 25 \kms\ for the whole
evolution.

The introduction of thermal conduction changes the details of the
evolution but not its main properties.  Figure 3 shows the case with
$\phi=0.3$; the results depend only weakly on the value of $\phi$ as
long as it is comparable to unity.  The snowplow and cooling mass
flows are contrasted by evaporation, which however becomes dominant
only after a few Myr.  The process of collapse of the diffuse phase is
completed in some 3.5 Myr, but after that time the fraction of diffuse
mass increases steadily.  Due to the higher density of the hot phase,
SNRs go to the PDS stage for the first 5.4 Myr, with a consequent
increase of energy losses.  The structure of the ISM is similar to
before, but, due to its higher density, the diffuse phase is
considerably colder and thermal pressure consequently lower.  Due to
the higher \ndif, SNRs are confined at a smaller radius and \fexit\
never reaches unit values; this implies that some fraction of the
kinetic energy is dissipated locally by turbulence.  The evolution of
the resulting SB is very similar, due to the weak dependence of its
expansion velocity on mechanical luminosity (Equation~\ref{eq:vsb}).

The system has been evolved with many combinations of initial
conditions and parameter values.  Figure 4 presents, as a function of
cloud mass, the fraction of diffuse ISM \fracd\ at 5 Myr, the
temperature \tdif\ of the diffuse phase at the same time, the time
\tdestr\ required by the SNe to complete the destruction of the cloud
(defined as the time at which the mass of the diffuse phase gets
smaller than 5 per cent more than its minimum) and the fraction
\flost\ of the total energy budget of SNe lost at 5 Myr.  The
quantities \fracd, \tdif\ and \flost\ are measured at 5 Myr as at this
time the destruction of the cloud is concluded in most cases; at later
times \fracd\ and \flost\ are constant in absence of
photo-evaporation, but increases otherwise, while \tdif\ increases.
However, it must be noticed that the size of the cloud at this point
is larger than the typical size of galaxy discs, so the applicability
of this model at late times is doubtful.

For the standard choice of parameters and initial conditions (as that
used in Figure 2), from 3 to 18 per cent of the mass ends up in the
diffuse phase, with \tdif\ ranging from $4\times10^5$ to $2.5\times10^6$
K; smaller clouds have lower \rsn\ values, so they maintain more
diffuse matter to a lower final temperature.  These two trends tend to
compensate each other, so the thermal energy contained in the diffuse
phase is roughly proportional to the cloud mass.  The time \tdestr\ is
\circa3 Myr and depends only weakly on the cloud mass, while the
fraction of energy lost by radiation is well below 10 per cent but for
the most massive clouds.

When evaporation is switched on, the amount of diffuse matter
increases significantly, especially for smaller cloud, while its
temperature decreases to $10^5$ K.  This trend is stronger for larger
$\phi$ values, although the dependence on the actual $\phi$ value is
weak. As a result of the lower temperature, the thermal energy of the
diffuse phase is lower than the no-evaporation case.  Destruction
times remain between 2 and 5 Myr, while radiation losses increase
but remain generally lower than 20 per cent.

The other panels show what happens for the non-evaporative case when
the parameter \fcool, the initial conditions (\fracd\ or \tdif) or the
external densities are varied.  In general, where more diffuse mass is
obtained its temperature is lower, so the thermal energy in the
diffuse phase does not vary as strongly as \fracd\ or \tdif.  The
destruction time in most cases ranges from 1 to 5 Myr, while the
fraction of energy lost is quite stable.  At variance with what found
in paper I, the results of the integration depend very sensitively on
the uncertain parameter \fcool, and thus on the detailed density
structure of the ISM.  Moreover, they depend significantly on the
state of the cloud when the SNe start to explode.  Besides, the
dependence on the external medium is not strong, though it is
important to notice that with a denser external ISM \fracd\ tends to
be lower and \tdif\ higher.

A detailed justification of all the trends visible in Figure 4 would
shed some light on the dynamics of the system, at the cost of a
lengthy discussion.  The most important information that this analysis
yields is a direct impression on the robustness of the predictions.

\begin{figure*}
\centerline{
\includegraphics[width=8.4cm,height=7cm]{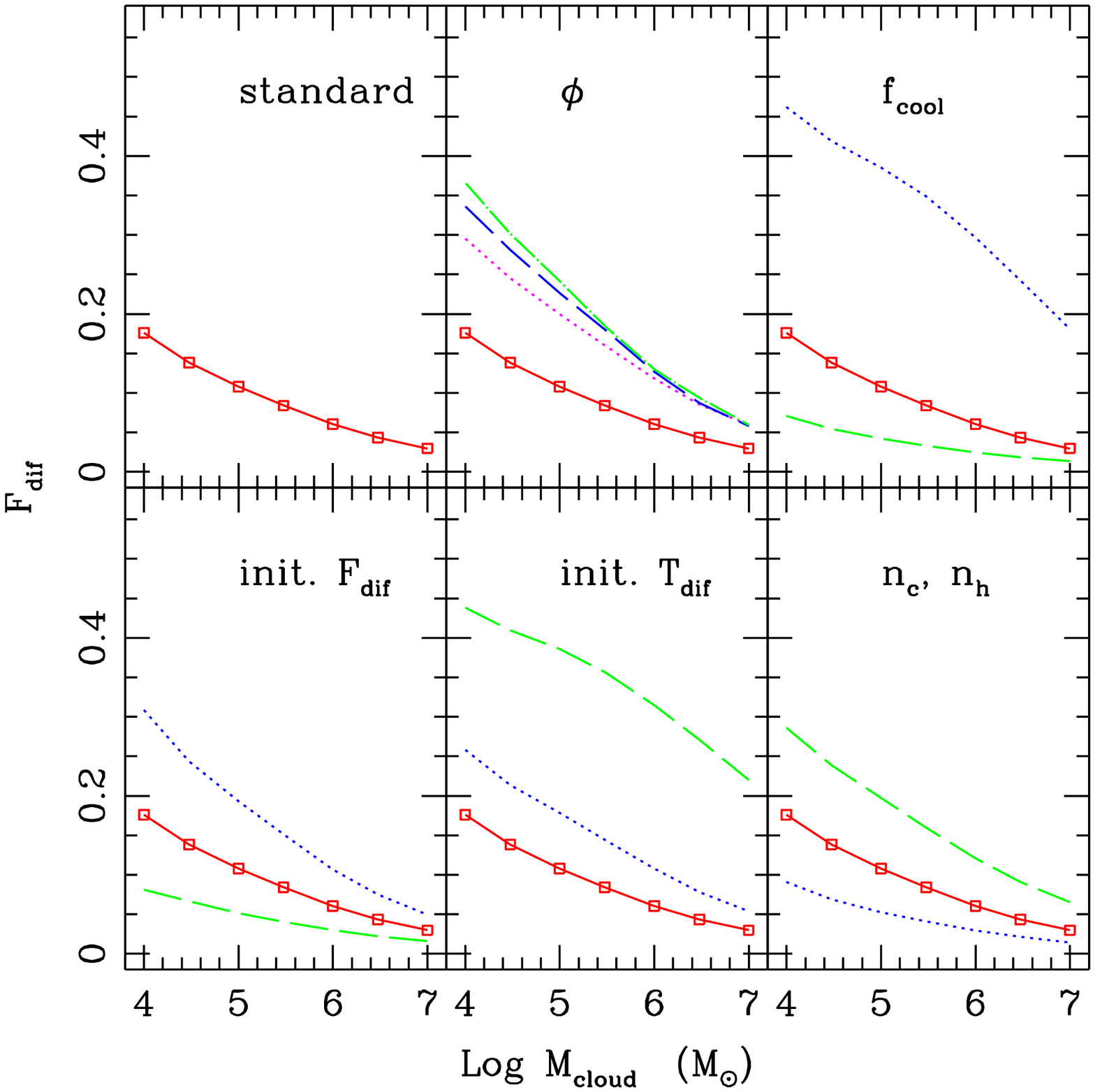}
\includegraphics[width=8.4cm,height=7cm]{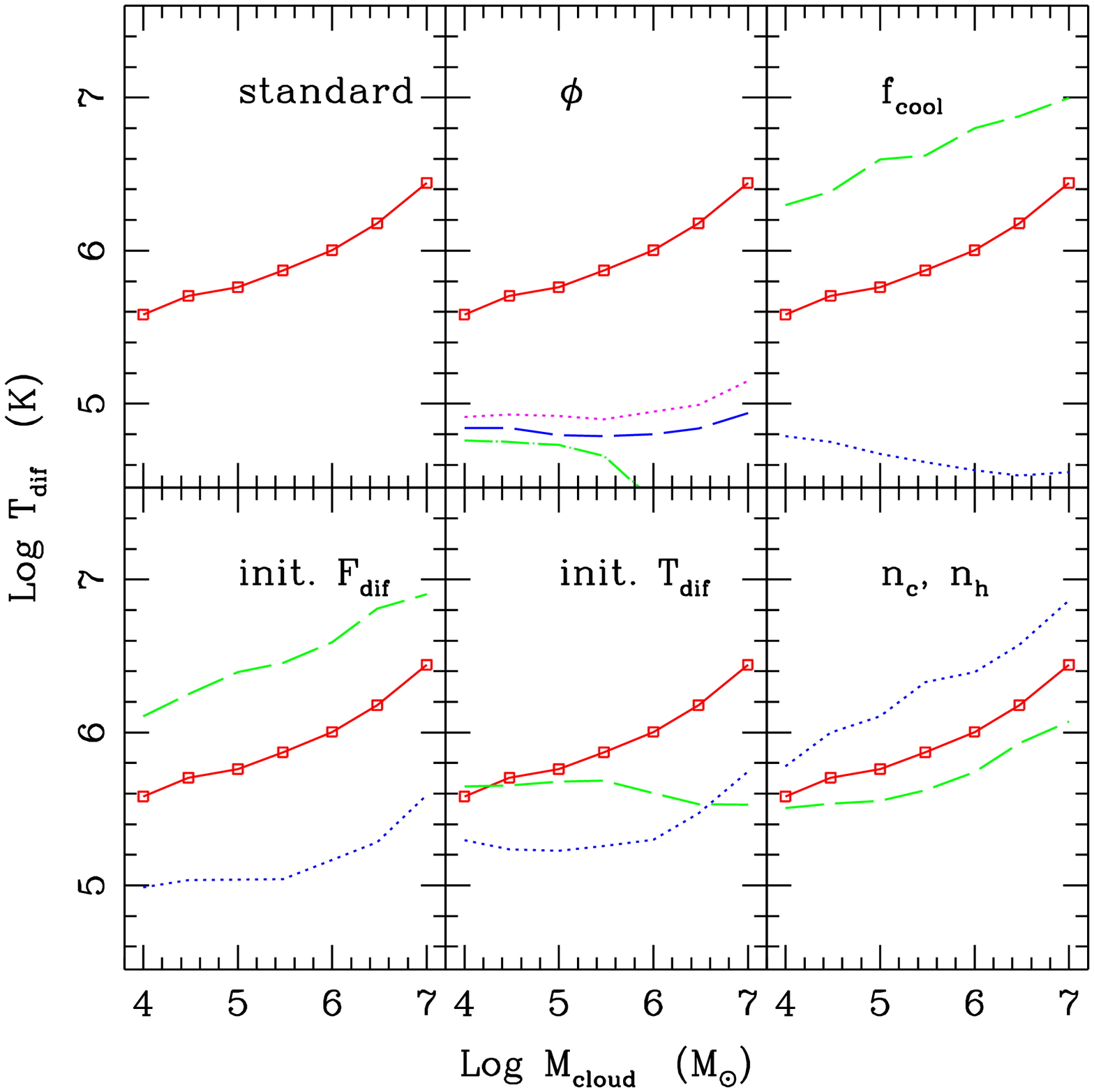}
}
\centerline{
\includegraphics[width=8.4cm,height=7cm]{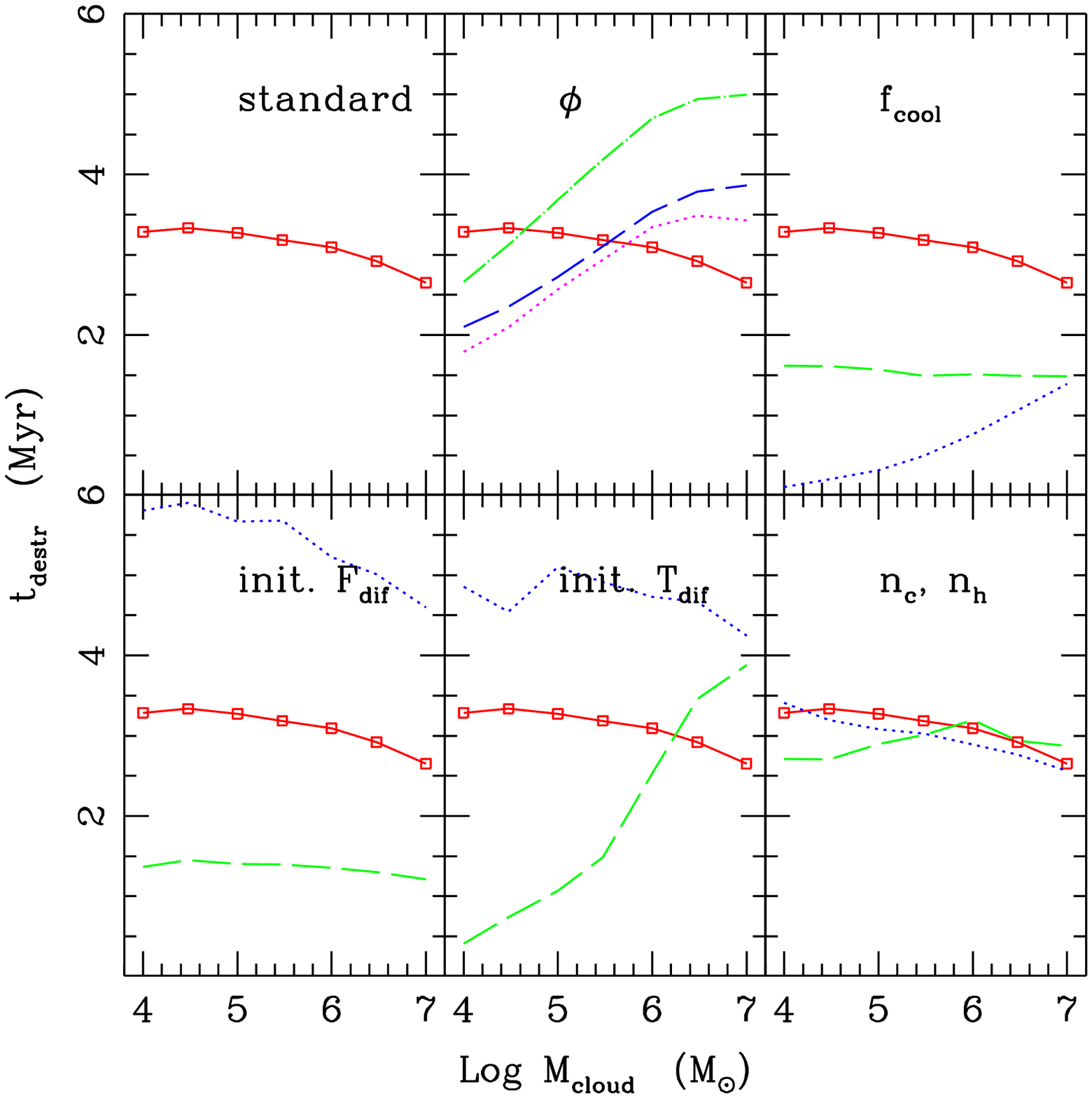}
\includegraphics[width=8.4cm,height=7cm]{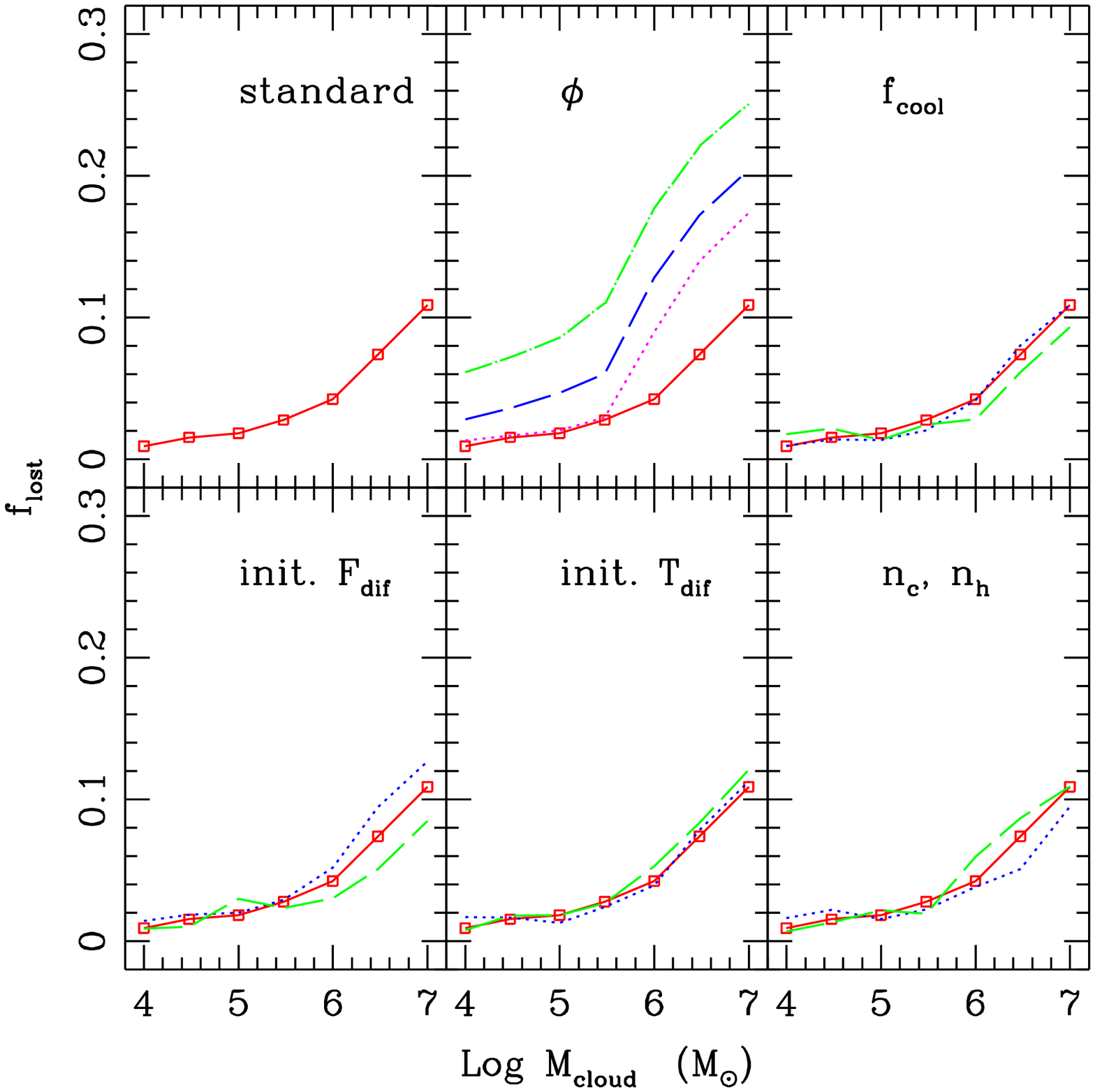}
}
\caption{\fracd, \tdif\ (both at 5 Myr), \tdestr\ and \flost\ for many
system as a function of cloud mass \mcloud.  Standard refers to the
choice of parameters of Figure 2, and is replicated in all panels.
$\phi$: dotted, dashed and dot-dashed lines refer to $\phi$=0.1, 0.3
and 1. \fcool: dotted and dashed lines refer to \fcool=0 and 0.3.
Initial \fracd: dotted and dashed lines refer to \fracd=0.9 and
\fracd=0.2.  Initial \tdif: dotted and dashed lines refer to
\tdif=$3\times10^4$ and $10^5$ K.  \nc, \nh: dotted line refers to
\nc=100, \nh=$10^{-2}$ \cmt, dashed line refers to \nc=1,
\nh=$10^{-4}$ \cmt.}
\label{fig:compl}
\end{figure*}

\section{Cloud destruction in the adiabatic confinement regime}

The system of Equations~\ref{eq:system} is valid as long as many SNe
concur in determining the evolution of the system.  More specifically,
many SNe must explode in the first 3-5 Myr, which implies (for
\tlife=27 Myr) a total number of SNe $\gg$10 and then a cloud mass
\mcloud$\gg$$10^4$ \msun.  For small clouds, the collapse of the
diffuse phase and the onset of adiabatic confinement before PDS take
place before the second SN manages to explode.  In this case the
present model simply does not apply.

As shown in paper I, such small collapsing clouds are not found in
spiral galaxies, where the Jeans mass (including non-sphericity and
turbulent support) is rather high.  This is a predicted result of the
feedback regime in which SBs blow out of the disc, injecting their
energy directly into the halo more than into the ISM.  If the system
is thicker or denser than a typical galaxy disc, SBs are kept
pressure-confined by the hot phase, so they inject all of their energy
into the ISM, which is then characterized by a much higher pressure
and temperature of the hot phase.  In this condition the Jeans mass is
considerably lower, and collapsing clouds range from 2000 to 10000
\msun.  As a consequence, only a bunch of SNe explode in each of them
(this is especially true if \fstar\ is set to a low value).

Let's consider the case of a \mcloud=2000 \msun\ uniform cloud with
density \nc\ as large as $10^3$ \cmt, and let's call
$M_{2000}=\Mcloud/2000$ \msun\ and $n_{1000}=\Nc/1000$ \cmt.  \rcloud,
computed from Equation~\ref{eq:fragmass}, will be

\be \Rcloud=2.5\ M_{2000}^{1/3} n_{1000}^{-1/3}\ {\rm pc}\, . \ee

\noindent
Neglecting for the moment photo-evaporation and any previous heating
by UV photons, the SNR will go into the PDS stage well before exiting
the cloud, as: 

\be  R_{\rm pds}=0.75\ n_{1000}^{-3/7}\ {\rm pc}\, .\ee

%

\noindent
The blast will reach the outer boundary of the cloud after

\be  t_{\rm out}=1.01\times 10^4\ M_{2000}^{1.11} n_{1000}^{-0.25}
\ {\rm yr}\, ,\ee

\noindent
with a final velocity of

\be  v_s=69\ M_{2000}^{-0.77} n_{1000}^{-0.08}\ {\rm km\ s}^{-1}\, ,\ee

\noindent
high enough not to be confined by the thermal or kinetic pressure of
the cloud.  At this point it will accelerate and fragment, spreading
around the mass of the star-forming cloud.  In this process it would
lose nearly all of its energy.  Any further SN would then explode in
the rarefied bubble, reaching easily the outer boundaries of the cloud
and merging with the external hot phase.  The fate of secondary SNe is
very difficult to predict, but a rule of thumb would suggest that
while the energy of the first SN is lost, a significant fraction of
the others will be injected into the medium.  In this case the
fraction of re-heated matter is likely to be rather low, in agreement
with the extrapolation of the results of Figure 4.

According to Matzner (2002), a single HII region can destroy a small
molecular cloud.  However, while UV photons can destroy all molecules
and make the cloud unbound, thus lowering the average density of the
cloud, it is likely that, for an external hot phase with $P/k\sim
10^5$ K cm$^{-3}$ and $T_{\rm h}\sim 10^7$ K, the region in which the
first SN explodes will be much denser then the external hot phase. The
SN will thus lose all of its energy and create a hot rarefied bubble
in place of an overdensity\footnote{This can be checked by considering
that $R_{\rm pds}/\Rcloud\propto n_{1000}^{-0.09}$ and $v_s\propto
n_{1000}^{-0.08}$.}.  So, the tentative conclusions given above remain
valid also in the case of pre-heated clouds.

\section{Discussion and conclusions}

Type II SNe are probably the most important source of stellar
feedback, but they explode in the densest regions of the ISM, the
star-forming molecular clouds.  The energetic input by expanding HII
regions and stellar winds can self-limit star formation to a low
efficiency and destroy the cloud in \circa10 Myr.  While the very
first SNe will likely remain trapped within the HII regions, the bulk
of them will explode when the cloud is already in the process of being
destroyed.

Under the assumption of a two-phase medium in pressure equilibrium,
the system evolves into a configuration where most mass is in a
collapsed, low filling-factor cold phase, while most volume is filled
by a hot pervasive phase, able to confine the expanding SNRs while
they are still in the adiabatic stage.  This way most energy is
efficiently pumped into the diffuse phase and used to power an SB
expanding in the external medium.  This is a very important point:
thanks to the multi-phase nature of the ISM, only a few per cent of
the energy injected by type II SNe is lost to radiation.

The destruction of the cloud is completed by SNe in \circa3 Myr, with
a loss of energy of \circa5-10 per cent, to which one might add the
energy of the very first SNe (roughly one per OB association) that are
kept trapped within the HII regions.  At the end of this process, a
fraction of mass of the original cloud, ranging from 5 to 30 per cent,
is heated to a temperature ranging from 10$^5$ to 10$^7$ K.  This last
quantity is especially uncertain; it depends sensitively on very
uncertain parameters like $\phi$, \fcool, on the specific initial
conditions and also on the way energy is given to the external ISM to
drive the SB.  On the other hand, the total thermal energy given to
the final diffuse phase is more robust.  Anyway, the choices used in
paper I of \fracd=0.1, \tdif=10$^6$ K and \flost\circa 0 are justified
by these results.

Excluding the very first explosions, SNe start to explode after a few
Myr, and complete the destruction of the cloud in \circa3 Myr.  In
this case, star formation cannot last more than several Myr.  The
destruction time inferred by Matzner (2002) of 10-30 Myr, based on the
role of HII regions alone, can then be overestimated by a factor
\circa2-3.  This would also apply to its predicted efficiency of star
formation of \circa10 per cent.  Low values of \fstar\circa 5 per cent
and of the duration of star formation, $\la$10 Myr, are indeed in good
agreement with observations (see, e.g, Carpenter 2000; Elmegreen
2000).

These numbers refer to relatively large star-forming clouds, expected
in galaxy discs where SBs are able to blow-out in the vertical
direction.  In this case, only 5-10 per cent of the energy budget of
SNe is injected into the ISM (paper I), while a similar amount is lost
in the destruction of the cloud.  The rest of the budget, \circa80 per
cent, is injected into the halo, and is thus available to heat up the
virialized halo gas, so as to preventing further cooling.

Smaller and denser clouds, predicted in paper I to be associated to
adiabatic confined SBs, are destroyed by one single SN.  In this case
the fraction of energy lost in the destruction is likely to be higher,
while the fraction of re-heated mass is likely to be lower.

Thermo-evaporation of cold clouds can play a significant role in this
context.  When $\phi$ is not negligible the diffuse mass at the final
time is more abundant and colder, and the energy loss is greater.
However, thermo-evaporation does not change the qualitative behaviour
of the system, and its contribution is still smaller than the
uncertainties connected to the values of the other parameters.  As a
consequence, the main contribution to the reheating of the cold phase
could be simply due to the residual effect of photo-evaporation by HII
regions.

While it is clear that only high-resolution MHD simulations of the
ISM, able to include all the relevant physical processes in play, will
be able to provide in the future robust quantitative predictions for
the destruction of a star-forming cloud, the calculations presented
here show that, thanks to the multi-phase nature of the ISM, a great
part of the energy from SNe will be able to leave the star-forming
regions, thus being available to regulate the formation of galaxies.

\section*{Acknowledgments}

The author thanks Chris Matzner for discussions.

{}

\bsp

\label{lastpage}


\begin{thebibliography}{}
\bibitem[]{ab} Ballesteros-Paredes, J., Vazquez-Semadeni, E., Scalo, J., 1999, ApJ 515, 286
\bibitem[]{mu} Carpenter, J.M., 2000, AJ, 120,3139
\bibitem[]{ai} Cioffi, D.F., McKee, C.F., Bertschinger, E., 1988, ApJ, 334, 252
\bibitem[]{el} Elmegreen, B.G., 2000, ApJ, 530, 277
\bibitem[]{ap} Elmegreen, B.G., 2002, ApJ, 564, 773
\bibitem[]{at} Franco, J., Shore, S.N., Tenorio-Tagle, G., 1994, ApJ, 436, 795
\bibitem[]{bc} Kritsuk, A., Norman, M.L., 2002, ApJ, 569, L127
\bibitem[]{bh} Mac Low, M.-M., Klessen, R.S., Burkert, A., Smith, M.D., 1998, PRL, 80, 2754
\bibitem[]{bk} Mac Low, M.-M., 2002, in ``Simulations of magnetohydrodynamic turbulence in astrophysics'', eds. T. Passot \& E. Falgarone, Springer.
\bibitem[]{ll} Mac Low, M.-M., 2003, in ``From Observations to Self-Consistent Modeling of the Interstellar Medium'', APSS, in press (astro-ph/0211616)
\bibitem[]{bm} Matzner, C.D., 2002, ApJ, 566, 30
\bibitem[]{mk} McKee, C.F., 1989, ApJ 345, 782
\bibitem[]{bn} McKee, C.F., Ostriker, J.P., 1977, ApJ, 218, 148
\bibitem[]{bo} McKee, C.F., van Buren, D., Lazareff, B., 1984, ApJ, 278, L115
\bibitem[]{br} Monaco, P., 2004, MNRAS, submitted (paper I)
\bibitem[]{bt} Ostriker, E.C., Gammie, C.F., Stone J.M., 1999, ApJ, 513, 259
\bibitem[]{bu} Ostriker, J., McKee, C.F., 1988, Rev.Mod.Phys., 60, 1
\bibitem[]{by} Press, W.H., Teukolsky, S.A., Vetterling, W.T., Flannery, B.P., 1992, Numerical recipes in FORTRAN, Cambridge: University Press
\bibitem[]{so} Solomon, P.M., Rivolo, A.R., Barrett, J., Yahil, A., 1987, ApJ, 319, 730
\bibitem[]{su} Sutherland, R.S., Dopita, M.A., 1993, ApJS, 88, 253
\bibitem[]{tm} Tan, J.C., McKee, C.F., 2004, in Formation and Evolution of Young Massive Clusters, eds. H. J. G. L. M. Lamers, A. Nota and L. Smith, in press (astro-ph/0403498)
\bibitem[]{th} Thornton, K., Gaudlitz, M., Janka, H.-Th., Steinmetz, M., 1998, ApJ, 500, 95
\bibitem[]{co} Vazquez-Semadeni, E., 2002, in ``Seeing Through the Dust: The Detection of HI and the Exploration of the ISM in Galaxies'', eds. R. Taylor, T. Landecker, \& A. Willis (ASP: San Francisco) (A larger version available in astro-ph/0201072)
\bibitem[]{cq} Weaver, R., McCray, R., Castor, J., Shapiro, P., Moore, R., 1977, ApJ, 218, 377
\bibitem[]{cs} Williams, J.P., McKee, C.F.,  1997, ApJ, 476, 166


\end{thebibliography}
\end{document}